\documentclass[aps,twocolumn,groupedaddress,superscriptaddress,floatfix,pra]{revtex4-1}
\usepackage{amsmath}
\usepackage{mathtools}
\usepackage{amssymb}
\usepackage{graphicx}
\usepackage{textcomp}
\usepackage{bm}	
\usepackage{soul}

\usepackage[usenames,dvipsnames]{color}

\usepackage[braket, qm]{qcircuit}

\begin{document}

\title{Maximally efficient quantum thermal machines fuelled by nonequilibrium steady states}


\author{Tiago F. F. Santos}
\affiliation{Instituto de F\'isica, Universidade Federal do Rio de Janeiro, CP68528, Rio de Janeiro, Rio de Janeiro 21941-972, Brazil}

\author{Francesco Tacchino}
\affiliation{IBM Quantum, IBM Research -- Zurich, CH-8803 R\"{u}schlikon, Switzerland}

\author{Dario Gerace}
\affiliation{Dipartimento di Fisica, Universit\`a di Pavia, via Bassi 6, I-27100, Pavia, Italy}
\author{Michele Campisi}
\affiliation{NEST, Istituto Nanoscienze-CNR and Scuola Normale Superiore
P.zza San Silvestro 12, I-56127 Pisa, Italy}
\affiliation{Dipartmento di Fisica e Astronomia, Universit\`a di Firenze, via Sansone 1, I-50019, Sesto Fiorentino (FI), Italy}
\affiliation{INFN Sezione di Pisa, Largo Pontecorvo 3, I-56127, Pisa, Italy}
\author{Marcelo F. Santos}\email{Corresponding author: mfsantos@if.ufrj.br}
\affiliation{Instituto de F\'isica, Universidade Federal do Rio de Janeiro, CP68528, Rio de Janeiro, Rio de Janeiro 21941-972, Brazil}

\pacs{xxxx, xxxx, xxxx}                                         
\date{\today}
\begin{abstract}
The concept of thermal machines has evolved from the canonical steam engine to the recently proposed nanoscopic quantum systems as working fluids. The latter obey quantum open system dynamics and frequently operate in non-equilibrium conditions. However, the role of this dynamics in the overall performance of quantum heat engines remains an open problem. Here, we analyse and optimize the efficiency and power output of two-stage quantum heat engines fuelled by non-equilibrium steady states. In a charging first stage, the quantum working fluid consisting of a qutrit or two coupled qubits is connected to two reservoirs at different temperatures, which establish a heat current that stores ergotropy in the system; the second stage comprises a coherent driving force that extracts work from the machine in finite a amount of time; finally, the external drive is switched off and the machine enters a new cycle.
\end{abstract}

\maketitle

\section{Introduction}
A heat engine is a device that produces work from heat, operating cyclically \cite{Callen}. A quantum heat engine is a heat engine whose working fluid is a quantum object, e.g., a few-level system \cite{ScovilDuBois, Kosloff2014, Kosloff1992}. Recent works have explored the differences and potential advantages of such quantum thermal machines when compared to their classical counterparts \cite{euro, prlPoem}, based on the possibility of exploiting genuinely quantum resources, such as coherence and quantum correlations \cite{prlPoem, Scully03Science299, LatuneSci2019, BarriosPRA2017,PhysRevLett.118.150601, Kamil2016, Scully2011, Goswami2013, Kammerlander2016}, or even the possibility to deliver alternative products, such as steady state entanglement \cite{Bellomo2013,BohrBrask2015,tacchino_steady_2018,Tavakoli2018}.

The study of quantum thermal machines can be traced back to the late Fifties,  when H.E.D.~Scovil and E.O.~Schulz-DuBois demonstrated that three-level masers can be treated as quantum heat engines \cite{ScovilDuBois}. Since then, many different quantum analogues of classical heat engines, operating equivalents of Carnot and Otto cycles \cite{Quan2007, Quan2006, Quan2005, Bender2000, Arnaud2002, friedemann2005, Kosloff2006, Kosloff2003, Thomas2011, Peterson2019, Popescu2014, Kosloff2015, Klaers2017}, as well as other general models  \cite{Alicki1979, Kosloff1984, Singer2016, Kosloff1996, Kosloff1994, Varinder2020, Boukobza2019, Sourav2020, Menczel2020, Gosh2018, Levy2016} have been presented. All these results have become particularly relevant in the last few years due to the development of new technologies at the nanoscale, where quantum effects become important. Furthermore, the field also presents fundamental theoretical challenges, such as describing the thermodynamics of measurements, the quantum version of fluctuations theorems, the above mentioned role played by coherence and entanglement, or the particularities of thermodynamics in discrete finite size systems \cite{Kammerlander2016, Pekola2015, Horodecki2013, Acin2013, Bejan1996, Suomela2016}.

Most studies in the field aim at establishing the operating limits and/or optimizing microscopic thermal machines working under realistic conditions \cite{Nori2007,Campisi2015,Kosloff2017,Landi2020}. In our previous work~\cite{MFS2020}, we have presented the principles of a newly designed quantum thermal machine fuelled by a non-equilibrium steady state (NESS) and operating with a two-stroke cycle composed of a non-unitary battery charging first stage and an ideal unitary work production second stage. We also exemplified its functioning by calculating the maximum efficiency and power output of a specific working fluid made of two weakly coupled qubits.\\
In the present work, we build on the former study by showing that the working fluid can be actually simplified down to a generic single three-level system (qutrit). This is, somewhat surprisingly, the simplest possible quantum system to produce work under the scheme previously introduced, and it allows envisioning an easier realization with state-of-art quantum technologies. We investigate the functioning of the quantum thermal machine under non-ideal conditions, thus establishing its overall operation limits and evaluating its performances by comparing the different working fluids that can be employed. More specifically, we consider finite time non-adiabatic work production stages for two distinct working fluids, namely the qutrit and the coupled qubits cases. For the latter, we revisit the quantum thermal machine working setup by relaxing the weak coupling condition that was previously considered~\cite{MFS2020}. We also compare different cycle periods and show that the asymptotic limit is reached when the charging period is taken to infinity. This calculation helps analysing the ideal limits of the quantum thermal machine, and the conditions to establish its cyclic operation.

The manuscript is organized as follows. We first revisit the theoretical principles underlying the general functioning of a NESS based quantum thermal machine under cyclic operation, in Sec.~\ref{sec:theory}. In Sec.~\ref{sec:results}, we introduce novel results for a specific implementation based on a single qutrit as the quantum working fluid, operated either in a $V$- or in a $\Lambda$-type energy level configuration; this is compared to the already introduced case of a working fluid made of a pair of coupled qubits, which is hereby described in its general configuration without limiting to weakly interacting qubits. Finally, in Sec.~\ref{sec:summary} we summarize and give concluding remarks and perspectives of the work. Extensive details on the calculations are reported in the Appendix.

\section{Theoretical Background}\label{sec:theory}

We hereby analyze the design principles and general operating performances of non-equilibrium quantum thermal machines. In our proposed framework, originally introduced within an idealized setup in Ref.~\cite{MFS2020}, these devices operate between two reservoirs at different temperatures, establishing a heat current through the working fluid made of a quantum system with well characterized energy spectrum. Such steady energy flow ``charges'' the system by leading it to an operational steady state $\rho_{OSS}$ that is ``active'' in the sense that it has positive ergotropy (namely one can withdraw energy from it by applying a unitary operation)
\cite{Ergotropy}. In the subsequent stroke, accordingly, a coherent drive is applied that withdraws the energy. 

In our description of the cycle the staring point of the cycle is the state $\rho_{OSS}$. In the first stage, an external drive is turned on for a finite time $\tau_d$, producing work and transforming the active state $\rho_{OSS}$ into a passive~\cite{PassiveStates} zero-ergotropy state $\rho(\tau_d)$. Then, in the second stage, the external drive is switched off and the heat current takes a time $\tau_r$ to restore the machine to $\rho_{OSS}$ leaving it ready for a new cycle. In the limit of large $\tau_r$ the operational steady state coincides with what is commonly referred to as the Non-equilibrium steady state, $\rho_{NESS}$.

We will assume a Markovian interaction between the quantum system and the heat reservoirs, such that the dynamics of both stages is governed by a master equation in the Lindblad form~\cite{OpenSystems} (In the following we shall adopt $\hbar = 1$ and $k_B = 1$):
\begin{equation}
\dot{\rho} = -i[H(t),\rho] + \mathcal{L}(\rho).
\label{eq1}
\end{equation}
The Hamiltonian reads $H(t) = H_0 + V(t)$, where $H_0$ is the free Hamiltonian of the system and $V(t)$ accounts for the coupling with the external work extraction drive (e.g., an electromagnetic field). The non-unitary part $\mathcal{L}(\rho)$ reads
\begin{equation}\label{dissipative}
    \mathcal{L}(\rho) = \sum_j L_j(\rho) =\sum_j  \Gamma_j\left[J_j \rho J_j^{\dagger} - \frac{1}{2}\{J_j^{\dagger}J_j, \rho \}\right],
\end{equation}
where $\{A,B\} = AB + BA$. $\Gamma_j$ are the transition rates and $J_j$ the respective jumps induced in the system by the heat reservoirs. In the recharging stage the external drive is switched off ($V(t) = 0$) and the energy gained during the recharging balances with the energy extracted from the system during the discharging, so that there is no variation in the internal energy of the system in a full cycle, i.e. $\Delta U_{cycle} = 0$.

At the end of the cycle, the total heat exchanged with the hot reservoir and the work performed by the machine are given by \cite{Alicki1979}
\begin{eqnarray}\label{genq}
    Q^H &=&  Q_r^H+Q_d^H \nonumber \\
    &=&\int_{0}^{\tau_d} dt \hskip 0.02cm \operatorname{Tr} [L_H(\rho(t))H(t)] \nonumber \\
    &+& \int_{\tau_d}^{\tau}dt \hskip 0.02cm \operatorname{Tr} [L_H(\rho(t))H_0],
\end{eqnarray}
\begin{equation}\label{genw}
    W = \int_{0}^{\tau_d} dt \hskip 0.02cm \operatorname{Tr} [\rho(t)\dot{H}(t)],
\end{equation}
where $L_H$ is the superoperator that represents the coupling to the hot reservoir, $\rho(t)$ is the density matrix of the working fluid, and $\tau = \tau_r + \tau_d$ is the total duration of the cycle. The efficiency of the machine is given by $\eta = -W/Q^H$, and its delivered power can be defined as $\mathcal{P} = -W/\tau$.

We remark that the analysis performed here significantly extends the scope and generality of previous works~\cite{MFS2020} by relaxing many simplifying assumptions and providing a compelling study of more realistic experimental conditions. In particular, Ref.~\cite{MFS2020} focused on the ideal scenario of an adiabatic discharging stage, where $\tau_d \ll \tau_r$ ($Q_d^H \rightarrow 0$). In such an approximation, the heat exchange only takes place in the charging stage, the period of the cycle becomes $\tau \approx \tau_r$, and the total heat exchanged between the system and the hot bath in a cycle approximates to
\begin{equation}\label{genqr}
    Q^H  \approx Q_r^H = \int_{0}^{\tau}dt \hskip 0.02cm \operatorname{Tr} [L_H(\rho^(t))H_0].
\end{equation}
Moreover, in the idealized scenario of Ref.~\cite{MFS2020}, the unitary evolution due to $V(t)$ was replaced by an ideal unitary operation, $U$, extracting the maximum amount of energy from $\rho_{OSS}$ and taking it to the completely passive state $\tilde{\rho}_{OSS} = U \rho_{OSS} U^{-1}$. In this case, the work generated in the cycle equals the full ergotropy of state $\rho_{OSS}$ given by~\cite{Ergotropy} 
\begin{equation}
\mathcal{E} = \sum_{k,j} r_k E_j (|\langle r_k|E_j \rangle|^2 - \delta_{kj}),
\end{equation}
where $E_j$ are the eigenenergies  of $H_0$ ordered in increasing magnitude, i.e $E_{i+1} > E_{i}$, and $r_k$ are the eigenvalues of $\rho_{OSS}$ ordered in decreasing magnitude, i.e.~$r_{k+1} < r_{k}$, with respective eigenvectors $\ket{E_j}$ and $\ket{r_j}$. Here, $U$ can be represented as $U = \sum_k \ket{E_k}\bra{r_k}$. This ideal situation produces a cycle of maximal efficiency $\eta_M = \mathcal{E}/Q_r^H$ and delivered power $\mathcal{P}_M = \mathcal{E}/\tau_r$. These values are the upper bounds and provide benchmarks for the more realistic scenarios that we analyse in this work, where for example the first stage takes finite time and cannot be considered fully adiabatic ($Q_d^H >0$). Such upper limit is achievable whenever $\tau_d \ll \Gamma_j^{-1}$, i.e. when the characteristic time of the discharging stage is much smaller than the inverse of the heat exchange rates. 

To summarize, at difference with previous studies, the framework introduced so far essentially solely relies on the Markovianity assumption of the system-reservoirs interaction, and remains fully general in all other aspects. In particular, it can be applied to any quantum system used as the working medium. In the following, we will concentrate our studies on the two simplest examples that represent minimal setups for the operation of a non-equilibrium quantum thermal machine between positive temperature reservoirs: a qutrit and two coupled qubits, respectively. 

\section{Results}\label{sec:results}

\subsection{A qutrit as the working fluid}

\begin{figure}
  \centering
   \includegraphics[width=\columnwidth]{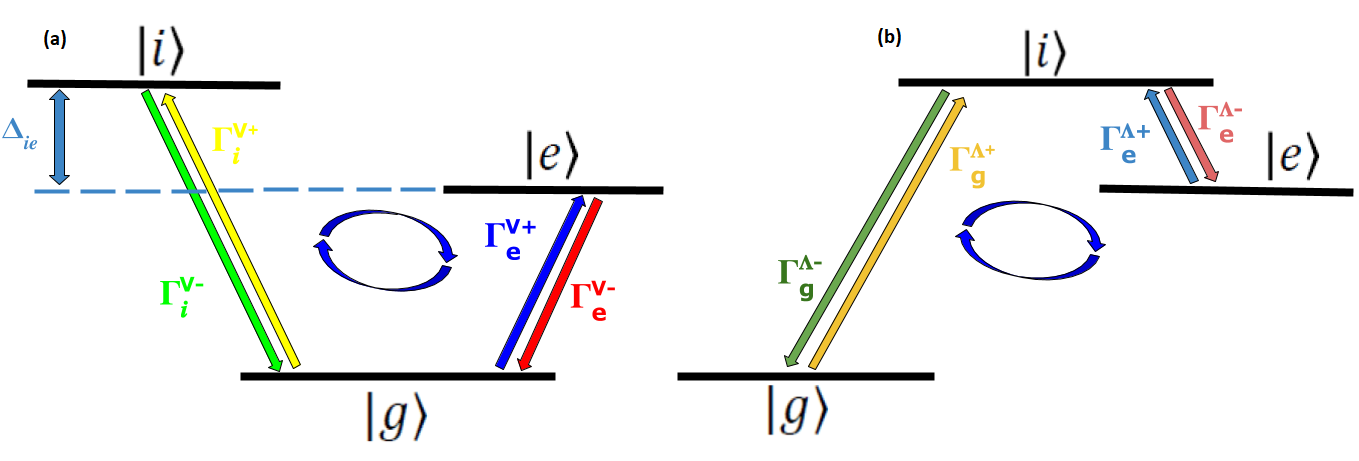}
   \caption{Configurations of the energies levels with transitions induced by the coupling with the heat baths for the qutrit as a quantum working fluid, in the V and $\Lambda$ configuration. Fig. (a) shows the qutrit in the V configuration. Here $\Delta_{ie} = E_i-E_e$. Fig. (b) shows the qutrit in the $\Lambda$ configuration.} \label{qutrit}
\end{figure}

We consider a single three-level system, a qutrit, as the simplest but nontrivial working fluid, as originally done in the pioneering work of Ref.~\cite{ScovilDuBois}. In particular, we focus on systems with internal levels $\{|g\rangle,|e\rangle,|i\rangle\}$ and respective energies $E_g = 0<E_e = \omega_e<E_i = \omega_i$ arranged in two different configurations: $V$ and $\Lambda$ (Fig.~\ref{qutrit}). In the $V$ configuration, the external heat sources couple the ground state, $|g\rangle$, to the excited states, $\ket{e}$ and $\ket{i}$, whereas in the $\Lambda$ configuration it is the excited state $|i\rangle$ that is coupled to the two lower energy states $|g\rangle$ and $|e\rangle$ by the heat reservoirs. In both cases, the third possible transition, namely $|e\rangle \rightarrow |i\rangle$ for $V$ and $|g\rangle \rightarrow |e\rangle$ for $\Lambda$, is not coupled through the heat reservoirs. 

Adapting the schemes previously presented~\cite{MFS2020}, a charging heat current can be created in the qutrit by coupling each transition to reservoirs kept at different temperatures. There are many different ways to engineer such reservoirs in distinct setups, such as superconducting or semiconducting qubits, trapped ions or atoms, among others \cite{Pekola2015,Josefsson2018,Rossnagel2016}. Here, we consider the generic case of a \textit{bona fide} heat bath and one or two auxiliary state selective incoherent sources.

The \textit{bona fide} heat bath at temperature $T$ generates Lindblad terms of the form:
\begin{eqnarray}\label{LV}
L_{j,k}(\rho)&=&\gamma_{j,k}(\bar{n}_{j,k}+1)[\sigma_{j,k}\rho \sigma_{k,j}-\frac{1}{2}\{\sigma_{k,k},\rho\}] \nonumber \\
&+& \gamma_{j,k}\bar{n}_{j,k}[\sigma_{k,j}\rho \sigma_{j,k}-\frac{1}{2}\{\sigma_{j,j},\rho\}],
\end{eqnarray}
where $\sigma_{j,k} = |j\rangle\langle k|$ ($\{j,k\}$ are the appropriate combinations of $\{g,e,i\}$), $E_k > E_j$ and $\bar{n}_{j,k} = (e^{\frac{E_k-E_j}{T}}-1)^{-1}$. Equation~\eqref{LV} alone produces a thermal (passive) steady state for which $\mathcal{E}=0$. To store ergotropy in the system, we need to add at least one extra reservoir, for example a heat source that incoherently pumps energy at rate $p$ from $\ket{g}$ to $\ket{i}$, and/or a heat dispenser that draws energy at rate $\Gamma$ from level $\ket{e}$ to level $\ket{g}$ (in the $V$ configuration) or from level $\ket{i}$ to level $\ket{e}$ (in the $\Lambda$ configuration). The overall effect of either one of this extra engineered reservoir is to take the system out of thermal equilibrium, creating the desired ergotropy in the working fluid. These extra reservoirs are described by the addition of new Lindblad terms to the dynamics of the qutrit: $L_p = p[\sigma_{i,g}\rho^j \sigma_{g,i}-\frac{1}{2}\{\sigma_{g,g},\rho\}]$ for the extra heat source and $L_{\Gamma}^{V}=\Gamma[\sigma_{g,e}\rho^{V} \sigma_{e,g}-\frac{1}{2}\{\sigma_{e,e},\rho^{V}\}]$ or $L_{\Gamma}^{\Lambda} = \Gamma \left[\sigma_{e,i} \rho^{\Lambda} \sigma_{i,e} - \frac{1}{2} \{\sigma_{i,i}, \rho^{\Lambda} \} \right]$ for the heat dispensers.

From the operational point of view, both cases can be studied under a common framework, where the dynamics of the thermal machine is given by Eq.~(\ref{eq1}), with $\mathcal{L}(\rho) = [L_{j,k}+L_{\Gamma}+L_p](\rho)$. In both configurations, the combination of these three reservoirs produces a dynamics that can be described by two effective temperatures, each one affecting one internal transition of the qutrit. For clarity, from now on, we treat each configuration separately.

\subsubsection{V configuration}
For the $V$ configuration, the non-unitary terms of Eq.(\ref{eq1}) can be rewritten as $\mathcal{L}(\rho) = \sum_k [L_k^{V+} + L_k^{V-}](\rho)$, where
\begin{equation}\label{LV+}
L_k^{V+}(\rho) = \Gamma_k^{V+} \left[\sigma_{k,g} \rho \sigma_{g,k} - \frac{1}{2} \{\sigma_{g,g}, \rho \} \right],
\end{equation}
\begin{equation}\label{LV-}
L_k^{V-}(\rho) = \Gamma_k^{V-} \left[\sigma_{g,k} \rho \sigma_{k,g} - \frac{1}{2} \{\sigma_{k,k}, \rho \} \right]
\end{equation}
and the transition rates are given by $\Gamma_e^{V+} = \gamma_e \bar{n}_e$, $\Gamma_e^{V-} = \gamma_e ( \bar{n}_e + 1) + \Gamma$, $\Gamma_i^{V+} = \gamma_i \bar{n}_i + p$ and $\Gamma_i^{V-} = \gamma_i (\bar{n}_i + 1)$. The effective temperatures of each engineered reservoir are then given by
\begin{equation}
    T_e^{V} =\frac{E_e}{\log{\left[\frac{\Gamma_e^{V-}}{\Gamma_e^{V+}}\right]}}
\end{equation}
and 
\begin{equation}
  T_i^{V} = \frac{E_i}{\log{\left[\frac{\Gamma_i^{V-}}{\Gamma_i^{V+}}\right]}},  
\end{equation}
where $T_i^{V}>0$ ($\Gamma_i^{V-} > \Gamma_i^{V+}$) is the temperature of the hot reservoir.

The machine is designed such that the state of the system after the recharging stage is diagonal in the energy eigenstates of the qutrit, i.e. $\rho(\tau_r)\equiv \rho_{OSS}= \sum_j p_j \ket{E_j}\bra{E_j}$. This state is active whenever, due to the action of the engineered baths,  $p_g > p_i > p_e$. In such cases, the ergotropy of $\rho_{OSS}$ reads $\mathcal{E}_V = \Delta_{ie} (p_i - p_e)$. In principle, there may exist more general cases for which, e.g., $p_i > p_e > p_g$, but those correspond to population inversion caused by coupling the system to negative temperature reservoirs \cite{tacchino_steady_2018} which are out of the scope of this paper.

In the limit of an adiabatic discharging stage, the unitary transformation that brings $\rho_{OSS}$ to the corresponding passive state $\tilde{\rho}_{OSS}$ is essentially a SWAP between levels $\ket{e}$ and $\ket{i}$.  In a more general scenario, this can be achieved by switching on an external drive of the type $V^{V}(t) = \epsilon\hskip 0.01cm \left(\ket{e}\bra{i} e^{i (\omega_i-\omega_e) t}  + \ket{i}\bra{e} e^{-i (\omega_i-\omega_e) t}\right)$ for a time $\tau_d = \pi/2\epsilon$. The adiabatic condition is approached for driving rate $\epsilon \gg \Gamma_j$.

The work done by the system, using Eq. \eqref{genw} with $V^V(t)$ is given by
\begin{eqnarray}\label{WV}
  W^V  &=& \int_0^{\tau_d} i \epsilon (\omega_i-\omega_e) \nonumber \\ &\times& \textrm{Tr}\left[\rho\left(e^{i(\omega_i-\omega_e)t}\ket{e}\bra{i}-e^{-i(\omega_i-\omega_e)t}\ket{i}\bra{e}\right)\right] dt, \nonumber \\
  W^V  &=& \int_0^{\tau_d} i \epsilon (\omega_i-\omega_e) [\varrho_{ie}(t) -\varrho_{ei}(t)] dt, \nonumber \\
\end{eqnarray}
where $\varrho_{jk} = \langle j|e^{iH_0 t} \rho e^{-i H_0 t}|k\rangle$. Whereas, the incoming heat from the hot bath on both stages is given by
\begin{eqnarray}\label{hv}
  Q_d^{HV} &=& \omega_i (\varrho_{ii}(\tau_d) - \varrho_{ii_{OSS}}) - \int_0^{\tau_d}  i \epsilon \omega_i [\varrho_{ie}(t) -\varrho_{ei}(t)] dt,\nonumber \\
Q_r^{HV} &=& \omega_i (\varrho_{ii_{OSS}} - \varrho_{ii}(\tau_d)).
\end{eqnarray}
Note that $\varrho_{kk} = \rho_{kk}$. The complete calculations for this configuration are presented in appendix A.

\subsubsection{$\Lambda$ configuration}
In the $\Lambda$ configuration, it is the excited state $|i\rangle$ that is coupled to the two lower energy states $|g\rangle$ and $|e\rangle$ by the heat reservoirs. Once again, the third possible transition, $|g\rangle \rightarrow |e\rangle$ in this case, is not coupled through the heat reservoirs. The Lindblad terms are given by
\begin{equation}\label{Llambda+}
L_k^{\Lambda+}(\rho) = \Gamma_k^{\Lambda+} \left[\sigma_{i,k} \rho \sigma_{k,i} - \frac{1}{2} \{\sigma_{k,k}, \rho \} \right] ,
\end{equation}
\begin{equation}\label{Llambda-}
L_k^{\Lambda-}(\rho) = \Gamma_k^{\Lambda-} \left[\sigma_{k,i} \rho \sigma_{i,k} - \frac{1}{2} \{\sigma_{i,i}, \rho \} \right],
\end{equation}
where $k=\{g,e\}$ with rates $\Gamma_g^{\Lambda+} = \gamma_g \bar{n}_g + p$, $\Gamma_g^{\Lambda-} = \gamma_g (\bar{n}_g + 1)$, $\Gamma_e^{\Lambda+} = \gamma_e \bar{n}_e $ and $\Gamma_e^{\Lambda-} = \gamma_e (\bar{n}_e + 1) + \Gamma$. Similarly, the effective temperatures of each engineered reservoir are given by $T_e^{\Lambda} = \frac{E_i - E_e}{\log{\left[\frac{\Gamma_e^{\Lambda-}}{ \Gamma_e^{\Lambda+}}\right]}}$ and $T_g^{\Lambda} = \frac{E_i}{\log{\left[\frac{\Gamma_g^{\Lambda-}}{\Gamma_g^{\Lambda+}}\right]}}$, where $T_g^{\Lambda}>0$ ($\Gamma_g^{\Lambda-} > \Gamma_g^{\Lambda+}$) is the temperature of the hot bath.

Similarly to the V configuration, the unitary transformation that brings $\rho_{OSS}$ to the corresponding passive state $\tilde{\rho}_{OSS}$ is a SWAP, this time between levels $\ket{g}$ and $\ket{e}$, and the corresponding coupling to a work extraction external drive is given by the Hamiltonian term $V^{\Lambda}(t) = \epsilon\hskip 0.1cm(\ket{e}\bra{g} e^{-i \omega_e t}  + \ket{g}\bra{e}e^{i \omega_e t} )$, turned on for a time $\tau_d = \pi/2\epsilon$. The work extracted and the heat gained from the hot source are given by
\begin{eqnarray}\label{WL}
W^{\Lambda} &=& \int_0^{\tau_d} i \epsilon \omega_e [\varrho_{eg}(t) -\varrho_{ge}(t)] dt,\nonumber \\
Q_d^{H\Lambda} &=&\int_0^{\tau_d}  i \epsilon \omega_i \left[\varrho_{ge}(t) -\varrho_{eg}(t)\right] dt  -\omega_i \left[\rho_{gg}(\tau_d) - \rho_{gg_{OSS}}\right],\nonumber \\
Q_r^{H\Lambda} &=& \omega_i \left[\varrho_{gg}(\tau_d) - \varrho_{gg_{OSS}}\right].
\end{eqnarray}
The complete calculations for the $\Lambda$ configuration are presented in appendix B.

\subsubsection{Overall results for the qutrit}
The works and heats expressed in Eqs.\eqref{WV}, \eqref{hv} and \eqref{WL} allow us to obtain the efficiencies for both configurations, which are given by
\begin{eqnarray}\label{nV}
\eta_V &=& \frac{-W^{V}}{Q_d^{HV} + Q_r^{HV}} = 1 - \frac{\omega_e}{\omega_i},\nonumber \\
\eta_\Lambda &=& \frac{-W^{\Lambda}}{Q_d^{H\Lambda} + Q_r^{H\Lambda}} = \frac{\omega_e}{\omega_i}.
\end{eqnarray}
Remember that $\omega_i>\omega_e$ by design, therefore, $\eta$ is limited to 1, as expected. Also notice that both efficiencies are maximized by increasing the energy gap between the two incoherent transitions of the system. This is achieved in both cases whenever the intermediate level $|e\rangle$ is almost degenerated with the energy level that is common to both reservoirs coupling, $| g\rangle$ for $V$, $|i\rangle$ for $\Lambda$. This gap is still limited by the fact that for Eq.~\ref{eq1} to hold, with the reservoirs given by Eq.~\ref{LV}, $\omega_e$ cannot approach zero in the $V$ configuration or $\omega_i$ in the $\Lambda$ configuration. This would invalidate the assumptions to derive the effect of the correspondent heat bathes on the particular transition (the $L_{g,e}$ term for $V$ and the $L_{e,i}$ term for $\Lambda$). This means that the machine efficiency can never actually be one. However, notice that it can be very close to the ideal limit. For example, for realistic atomic or ionic working fluids, one transition can involve the exchange of optical photons, while the other can operate with microwave ones. Therefore, the ratio $\omega_e/\omega_i$ ($(\omega_i-\omega_e)/\omega_i$) in the $V$ ($\Lambda$) configuration can easily be as small as $10^{-4}$.

Another interesting property is that when the efficiency in $V$ scheme is maximized, the efficiency in $\Lambda$ configuration is very small, and vice-versa. Also notice that both efficiencies only depend on the unperturbed energy spectrum of the system, meaning that, for qutrits in $V$ or $\Lambda$ configurations as working fluids and for fixed energy levels, efficiency is constant no matter how fast the battery is charged and discharged. In fact, these are the efficiencies achieved when the discharging stage is adiabatic. Finally, we notice that this efficiency is still limited by the Carnot efficiency of an equivalent machine operating under the same temperature gradient. For example, in the $V$ configuration, the equivalent Carnot efficiency is given by $\eta_{Carnot}=1-\frac{T_e}{T_i}$. For the machine to produce work, the state of the working fluid needs to be active at the beginning of the cycle, i.e. one needs $\mathcal{E}>0$. This requires $p_i>p_e$ which is achieved whenever $\frac{\Gamma_i^{V+}}{\Gamma_i^{V-}} >\frac{\Gamma_e^{V+}}{\Gamma_e^{V-}}$. This implies that $\omega_e/\omega_i>T_e/T_i$, and, therefore, $\eta_V<\eta_{Carnot}$. A similar calculation holds for the system in $\Lambda$.

As an example, we compute the efficiency for the $V$ configuration operating in an ideal short cycle (SC), i.e., when $\epsilon \gg \Gamma_j$ (adiabatic work extraction) and $\sum_j \Gamma_j \tau \ll 1$. We recall that these represent sufficient conditions for optimal efficiency and output power for weakly coupled qubits~\cite{MFS2020}. In this limit, the ergotropy stored in $\rho_{OSS}^{V}$ becomes proportional to the cycle duration $\tau \sim \tau_r$, and it is given by 
\begin{equation}
    \mathcal{E}_{SC}^V = (\omega_i - \omega_e) K_V \tau,
\end{equation}
where
\begin{equation}
   K_V = \frac{\Gamma_e^{V-}\Gamma_i^{V+} - \Gamma_e^{V+}\Gamma_i^{V-}}{2 (\Gamma_e^{V+} + \Gamma_i^{V+}) + \Gamma_e^{V-} + \Gamma_i^{V-}}.
\end{equation}
At the same time, the heat exchanged with the hot bath at effective temperature $T_i^{V}$ is also proportional to $\tau$ and given by
\begin{equation}
    Q_{SC}^{HV} = \omega_i K_V \tau,
\end{equation}
resulting in a machine of efficiency $\eta_{SC}^V  = 1 - \omega_e/\omega_i$, as expected. The corresponding result for the $\Lambda$ configuration can be recovered with similar calculations.

The generated power $\mathcal{P}_{SC}^j  = \mathcal{E}/\tau$ ($j=V,\Lambda$) reads $\mathcal{P}_{SC}^V=(\omega_i - \omega_e) K_V $ and $\mathcal{P}_{SC}^\Lambda= \omega_e K_\Lambda$, where 
\begin{equation}
K_\Lambda = \frac{\Gamma_e^{\Lambda-}\Gamma_g^{\Lambda+} - \Gamma_e^{\Lambda+}\Gamma_g^{\Lambda-}}{2 (\Gamma_e^{\Lambda-} + \Gamma_g^{\Lambda-}) + \Gamma_e^{\Lambda+} + \Gamma_g^{\Lambda+}}    
\end{equation}
Notice that, differently from the efficiencies, the output powers depend not only on the respective spectra, but also on heat current rates $\Gamma_k^{j\pm}$, where $k = e,i$ for $j = V$ and $k = g,e$ for $j = \Lambda$. The most powerful configuration will depend on the specific details of the heat flow and the energy levels of the working fluid. An interesting comparison can be done, for example, by considering fixed and equal spectra for both configurations, the same vacuum coupling to each reservoir $\gamma_i = \gamma_e = \gamma_g = \gamma$, $\Gamma \rightarrow 0$ and $T \rightarrow 0$. In this case, we obtain \begin{equation}
\lim_{T, \Gamma \to 0} \frac{\mathcal{P}_{SC}^V}{\mathcal{P}_{SC}^{\Lambda}} = \frac{\omega_i}{\omega_e} - 1    
\end{equation}
i.e.~the $V$ configuration delivers more power if $\omega_i/\omega_e > 2$, while the $\Lambda$ configuration maximises the power output if $1 < \omega_i/\omega_e  < 2$.

\subsection{Two coupled qubits as the working fluid}

We now turn our attention to a working fluid composed of two coupled qubits~\cite{MFS2020}, described by the Hamiltonian
\begin{equation}
H_0 = \omega_0 (\sigma_+^{(1)}\sigma_-^{(1)} + \sigma_+^{(2)}\sigma_-^{(2)}) + \lambda (\sigma_+^{(1)}\sigma_-^{(2)} + h.c)
\end{equation}
where $2\sigma_{\pm}^{j} = \sigma_x^j \pm i\sigma_y^j$, $j = 1,2$, and $\{\sigma_k\}$ for $k = x, y, z$ are the Pauli matrices. The heat current is established by coupling the system to two reservoirs of temperatures $\mathcal{T}_A>\mathcal{T}_S$ that act respectively on its symmetric (S) and anti-symmetric (A) subspaces. The energy spectrum has four levels, $\{|G\rangle, |S\rangle,|A\rangle,|E\rangle\}$, already ordered by increasing energy $\{E_{G},E_S,E_A,E_{E}\}$, where $|G\rangle=|gg\rangle$,  $|S\rangle = \frac{|ge\rangle+|eg\rangle}{\sqrt{2}}$, $|A\rangle = \frac{|ge\rangle-|eg\rangle}{\sqrt{2}}$ and $|E\rangle=|ee\rangle$. In the weak coupling limit ($\lambda \ll \omega_0$), levels $|S\rangle$ and $|A\rangle$ are close to each other and their energy distance to levels $|G\rangle$ and $|E\rangle$ are approximately the same. These conditions would allow us to assume, as done in Ref.~\cite{MFS2020}, that the bare coupling rates $\gamma_j$ of each non-unitary transition are approximately the same and the problem is restricted to four overall transition rates given by $\Gamma_{A, S}^{+} = \gamma_0 \bar{n}_{A,S}$ and $\Gamma_{A, S}^{-} = \gamma_0 (\bar{n}_{A,S} + 1)$, where $\bar{n}_{A,S} = (e^{\frac{\omega_0}{T_{A,S}}} - 1)^{-1}$. However, in the more general scenario that we will consider here, each coupling rate may depend on the energy gap of the respective transition. In such case, the heat current is established by combining eight different non-unitary channels, two for each one of the four incoherently coupled transitions.

\begin{figure}
  \centering
   \includegraphics[width=0.8\columnwidth]{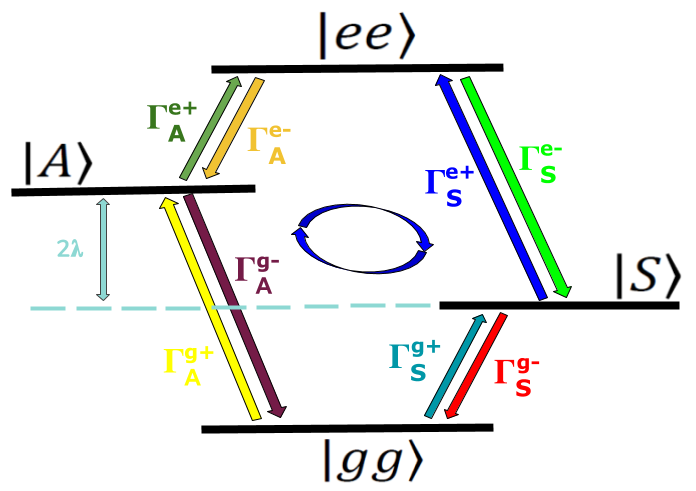}
   \caption{Level scheme for the two qubits coupled as the working fluid and the transition rates induced by the coupling with the heat baths. Here $E_A - E_S = 2 \lambda$.} \label{2qb}
\end{figure}

Similar to the qutrit case, the steady state $\rho_{NESS}$ is diagonal in the total spin basis, i.e,  $\rho_{NESS} = \sum_{i} r_{i_{NESS}} \ket{i}\bra{i} $, where $i = \{G,S,A,E\}$. However, different from the previous working fluid, $\rho_{NESS}$ can in general be entangled~\cite{tacchino_steady_2018}. Whenever $r_{A_{NESS}} > r_{S_{NESS}}$, the ergotropy, $\mathcal{E}_{NESS}$, stored in $\rho_{NESS}$ is given by (see Appendix C)
\begin{equation}
   \mathcal{E}_{NESS} = 2 \lambda ( r_{A_{NESS}} - r_{S_{NESS}})
\end{equation}
Once again, the machine can operate in cycles of arbitrary periods $\tau$. The operational steady states $\rho_{OSS} = \sum_{i} r_{i} \ket{i}\bra{i}$ will also be diagonal in the $i = \{G,S,A,E\}$ basis, and the ergotropy stored in $\rho_{OSS}$ will also be given by
\begin{equation}
    \mathcal{E}_{\tau} = 2 \lambda (r_A - r_S).
\end{equation}
However, in general, $\mathcal{E}_{\tau} \leq \mathcal{E}_{NESS}$, equality being reached for $\tau \rightarrow \infty$. Similarly to the qutrit case, in the ideal scenario of an adiabatic first stage (discharging), the unitary operation $U_{\mathcal{E}}$ that extracts the maximum amount of energy stored in the system is formally a SWAP between the $\ket{A}$ and $\ket{S}$ populations. If we now consider a realistic external drive producing such unitary, $V(t) = \epsilon \left(\ket{A}\bra{S} e^{i 2 \lambda t} + \ket{S}\bra{A} e^{-i 2 \lambda t}\right)$, where $E_A = \omega_0 + \lambda$ and $E_S = \omega_0 - \lambda$, the work done by the system is given by 

\begin{equation} 
W = 2 i  \epsilon \lambda \int_0^{\tau_d} [\varrho_{AS}(t) - \varrho_{SA}(t)] dt.
\end{equation}
If $\epsilon >> \Gamma_{A,S}^{k \pm}$, $k = G, E$, we obtain $W = - \mathcal{E}$ as expected (see appendix C). The incoming heat from the hot bath $\mathcal{T}_A$ during the discharging is given by
\begin{align}
Q_d^H &= \int_0^{\tau_d} 4 \omega_0[\Gamma_A^{E+} \varrho_{A}(t) - \Gamma_A^{E-} \varrho_{E}(t)] dt  \nonumber \\
&+ i \epsilon (\omega_0 + \lambda) \int_0^{\tau_d} [\varrho_{SA}(t)-\varrho_{AS}(t)] \nonumber \\
&- (\omega_0 + \lambda) [\varrho_{A_{OSS}} - \varrho_A(\tau_d)] 
\end{align}
During recharging, there is no work performed by or on the system and the heat exchanged between the system and the hot bath is given by
\begin{align}
   Q_r^H &= \int_{\tau_d}^{\tau} 4 \omega_0 [\Gamma_A^{E+} \varrho_A(t) - \Gamma_A^{E-} \varrho_{E}(t)]dt \nonumber \\
   &+ (\omega_0 + \lambda)[\varrho_{A_{OSS}} - \varrho_A(\tau_d)],
\end{align}
This results in an efficiency of the machine given by
\begin{equation}\label{geneff}
\eta = \frac{-W}{Q_d^H + Q_r^H} = \frac{1 - \frac{E_S}{E_A}}{1 + \alpha + \beta}
 \end{equation}
where,
\begin{eqnarray}
\label{alpha2q1}
\alpha &=& \frac{\int_0^{\tau_d} 4 \omega_0 [\Gamma_A^{E+} \varrho_A(t) - \Gamma_A^{E-} \varrho_{E}(t)] dt}{i\epsilon (\omega_0 + \lambda) \int_0^{\tau_d} dt [\varrho_{SA}(t)-\varrho_{AS}(t)]} \nonumber \\
\beta &=& \frac{\int_{\tau_d}^{\tau} 4 \omega_0[\Gamma_A^{E+} \varrho_A(t) - \Gamma_A^{E-} \varrho_{E}(t)] dt }{i\epsilon (\omega_0 + \lambda) \int_0^{\tau_d} dt [\varrho_{SA}(t)-\varrho_{AS}(t)]}.
\end{eqnarray}
Notice that, contrary to the qutrit case, the existence of two (and not one) energy levels ($|G\rangle$ and $|E\rangle$) incoherently coupled to the work producing subspace $\{|A\rangle,|S\rangle\}$ results, in general, in a larger amount of incoming heat, lowering the efficiency of the machine. Here, the best efficiencies are reached when $\alpha+\beta$ is minimized. Since the explicit dependence of $\alpha$ and $\beta$ on the parameters $\lambda$, $\epsilon$ and $\Gamma_j$ involves a rather technical calculation, we will only present below some numerical results for a particular example that already encompass all the relevant features of the general case. Before going there, however, we can still gain some useful insight by an analytical treatment of the short cycle (SC) regime, where the expressions simplify significantly. Gladly, as demonstrated in Ref.~\cite{MFS2020}, this corresponds to the most interesting limit for the two-qubit working fluid, namely the one that produces higher efficiency and output power.

In SC conditions, the ergotropy stored in $\rho_{OSS}$ is, up to first order in $\tau$, given by 
\begin{equation}
\mathcal{E}_{SC} = 4 \lambda \kappa\tau,
\end{equation}
where $\Gamma_k^{\pm} = \Gamma_A^{k\pm} + \Gamma_S^{k\pm}$ and
\begin{equation}
    \kappa = \frac{\Gamma_E^{-}(\Gamma_A^{G+} \Gamma_S^{G-} - \Gamma_A^{G-} \Gamma_S^{G+}) \nonumber - \Gamma_G^{+}(\Gamma_A^{E+} \Gamma_S^{E-} - \Gamma_A^{E-} \Gamma_S^{E+})}{\Gamma_E^{-}(\Gamma_G^{+} + \Gamma_G^{-})
    + \Gamma_G^{+}(\Gamma_E^{+} + \Gamma_E^{-})}.
\end{equation}
The heat absorbed from the hot bath $\mathcal{T}_A$ is given by
\begin{eqnarray}
Q^H_{SC} &= \frac{2}{\Omega}[(\omega_0 - \lambda) \Gamma_G^+ (\Gamma_A^{E+}\Gamma_S^{E-} - \Gamma_S^{E+}\Gamma_A^{E-}) \nonumber \\
&+ (\omega_0 + \lambda) \Gamma_E^- (\Gamma_A^{G+}\Gamma_S^{G-} - \Gamma_S^{G+}\Gamma_A^{G-})]\tau,
\end{eqnarray}
where
\begin{equation}
    \Omega = \Gamma_E^{-}(\Gamma_G^{+} + \Gamma_G^{-})
    + \Gamma_G^{+}(\Gamma_E^{+} + \Gamma_E^{-}). \nonumber
\end{equation}
The efficiency of the machine operating in the ideal short cycle, $\eta_{SC}$, is given by
\begin{equation}\label{effsc2q}
    \eta_{SC}  = \left(1-\frac{E_S}{E_A}\right) \frac{1 - f}{1 + \frac{E_S}{E_A}f}
\end{equation}
where
\begin{equation}
    f = \frac{\Gamma_G^+(\Gamma_A^{E+}\Gamma_S^{E-} - \Gamma_S^{E+}\Gamma_A^{E-})}{\Gamma_E^- (\Gamma_A^{G+}\Gamma_S^{G-} - \Gamma_S^{G+}\Gamma_A^{G-})}.
\end{equation}
For a fixed $\lambda/\omega_0$ ratio, $\eta_{SC}$  is maximized when $f << 1$. This condition is reached when $\Gamma_A^{k-} + \Gamma_S^{k-} >> \Gamma_A^{k+} + \Gamma_S^{k+}$ which can be achieved either if $\Gamma_j^{k-} >> \Gamma_j^{k+}$ or if $\Gamma_S^{k-} >> \Gamma_A^{k-}$. In both cases, the maximum efficiency tends to
\begin{equation}\label{efmaxsc}
    \eta_{SC}^{max} \rightarrow 1-\frac{E_S}{E_A},
\end{equation}
which is essentially the same obtained for the qutrit, except that, for coupled qubits, it is only achievable in the short cycle. Note that this is also the efficiency of the short cycle of the weak coupling regime, although here there is no pre-determined restriction over the ratio $E_S/E_A$, that can, in principle, be made very close to 0, taking $\eta_{SC}^{max} \rightarrow 1$.

From the perspective of the generated power, in the first case, $\Gamma_j^{k-} >> \Gamma_j^{k+}$, the power in the short cycle depends on both temperatures and is given by $\mathcal{P}_{SC}^{max} \approx 2 (E_A - ES) \frac{\Gamma_A^{G+}\Gamma_S^{G-} - \Gamma_S^{G+}\Gamma_A^{G-}}{\Gamma_G^-}$ while, in the second case, it depends only on the temperature of the hot reservoir, $\mathcal{T}_A$, and it amounts to $\mathcal{P}_{SC}^{max} = 2(E_A-E_S) \Gamma_A^{G+}$.

\begin{figure*}[t]
  \centering
   \includegraphics[width=0.9\textwidth]{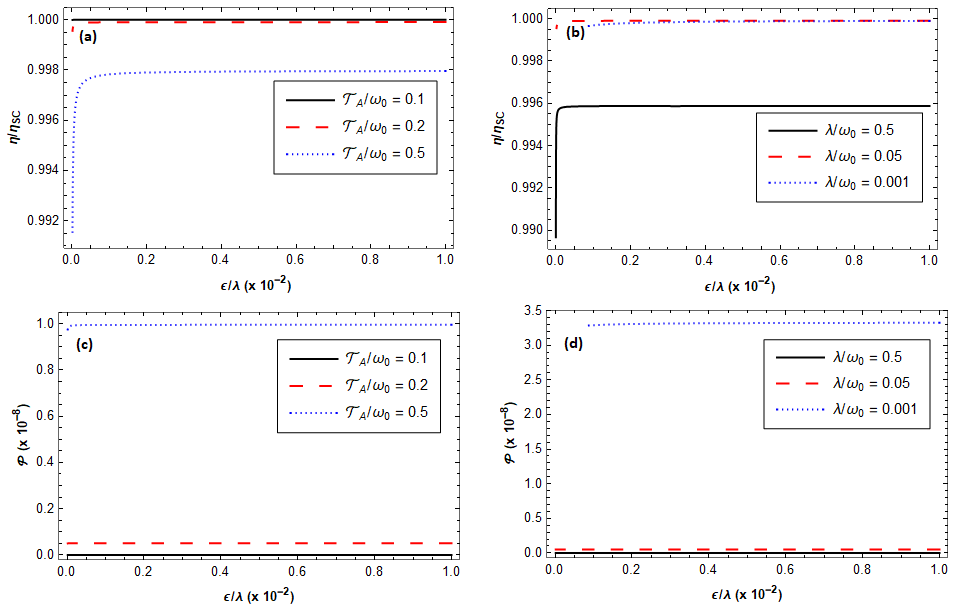}
   \caption{We plot $\eta/\eta_{max}$ and $\mathcal{P}$ as a function of $\epsilon/\lambda$ for the machine operating in the short cycle limit. Here $\eta_{SC}$ and $\eta$ are given by \eqref{geneff} and \eqref{effsc2q}, respectively, and $\mathcal{P} = -W/\tau$. Here we use $\gamma_0/\omega_0 = 10^{-8}$ and $1/\tau\omega_0 = 10^{-6}$. In figures (a) and (c) ((b) and (d)) we use $\mathcal{T}_A/\omega_0 = 0.2$ and $\mathcal{T}_S/\omega_0 = 0.05$ ($\lambda/\omega_0 = \mathcal{T}_S/\omega_0 = 0.05$).} \label{shortcycle}
\end{figure*}

Before concluding, let us now go back to an arbitrary cycle duration and calculate numerically the efficiency for different temperatures of the reservoirs and as a function of the strength of the work producing drive. We assume a radiative decay model for the coupling to the reservoirs (Fig. \ref{2qb}) where the transition rates are given by  $\Gamma_{A, S}^{E+} = \gamma_{0A, S}^{E} \bar{n}_{A, S}^E$, $\Gamma_{A, S}^{G+} = \gamma_{0A, S}^{G} \bar{n}_{A, S}^G$, $\Gamma_{A, S}^{E-} = \gamma_{0A, S}^{E} (\bar{n}_{A, S}^E + 1)$ and $\Gamma_{A, S}^{G-} = \gamma_{0A, S}^{G} (\bar{n}_{A, S}^G + 1)$. The average number of excitations in each reservoir is given by
\begin{eqnarray}
\bar{n}_A^E = \frac{1}{e^{\frac{\omega_0-\lambda}{\mathcal{T}_A}}-1}, \qquad & \bar{n}_S^E = \frac{1}{e^{\frac{\omega_0+\lambda}{\mathcal{T}_S}}-1} \\
\bar{n}_A^G = \frac{1}{e^{\frac{\omega_0+\lambda}{\mathcal{T}_A}}-1}, \qquad & \bar{n}_S^G = \frac{1}{e^{\frac{\omega_0-\lambda}{\mathcal{T}_S}}-1},
\end{eqnarray}
and the dependence of the bare coupling rates with the energy of the levels is $\gamma_{0S}^G = \gamma_{0A}^E = \gamma_0 (1 - \frac{\lambda}{\omega_0})^3$ and $\gamma_{0A}^G = \gamma_{0S}^E = \gamma_0 (1 + \frac{\lambda}{\omega_0})^3$. Notice that, as mentioned before, in the weak coupling limit ($\lambda \ll \omega_0$) all the bare rates are approximately the same. Whereas, on the other extreme of large coupling, we still want to restrict ourselves to the level configuration of Fig.~\ref{2qb} as well as, similar to the qutrit scenario, guarantee that the dynamics is given by Eq.~\ref{eq1} with the adequate non-unitary terms. In other words, $\lambda$ can approach $\omega_0$ but it has to be sufficiently smaller than the qubits bare frequency for us to remain far enough from level crossings. 

Finally, we will focus on the two most significant regimes of operation for the machine: the short cycle (SC), characterized by $\sum_j \Gamma_{A,S}^{k\pm} \tau \ll 1$, in which the stored ergotropy is proportional to the cycle duration and, on the other extreme, the asymptotic cycle (AC), in which the system is charged to its maximum possible ergotropy and the initial state $\rho_{OSS}$ converges to the {\it bona fide} steady state of Eq.(\ref{eq1}), $\rho_{NESS}$. In both cases we assume that the discharging stage is not adiabatic in the thermodynamic sense and its duration is chosen so that all the ergotropy stored in the system is extracted in the form of work. Efficiencies will be compared to the benchmark results obtained in Eq.~\eqref{effsc2q}.

\begin{figure*}[t]
  \centering
   \includegraphics[width=0.9\textwidth]{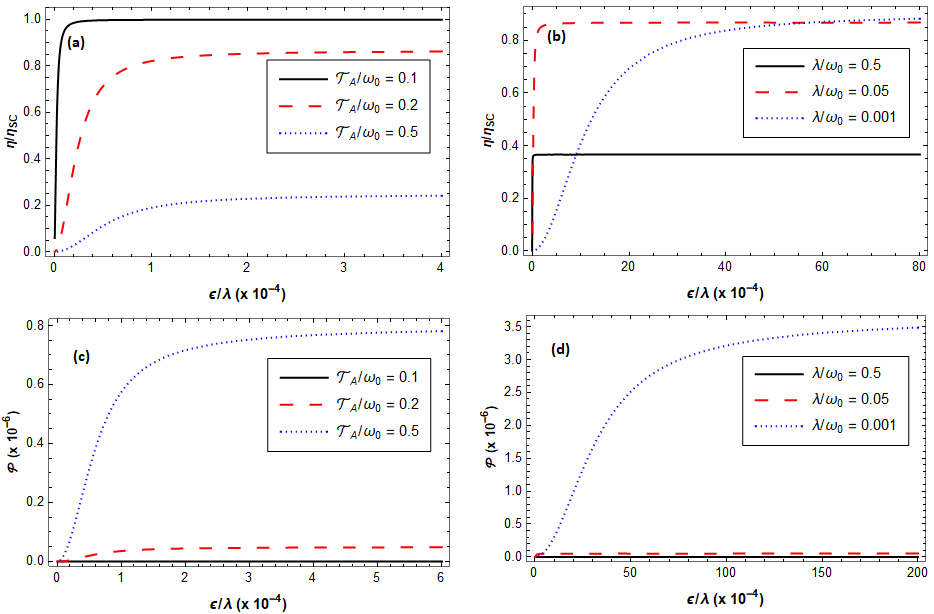}
   \caption{We plot $\eta/\eta_{max}$ and $\mathcal{P}$ as a function of $\epsilon/\lambda$ for the machine operating in a long cycle. Here $\eta_{SC}$ and $\eta$ are given by \eqref{geneff} and \eqref{effsc2q}, respectively, and $\mathcal{P} = -W/\tau$. Here we use $\gamma_0/\omega_0 = 10^{-5}$. In figures (a) and (c) ((b) and (d)) we use $\mathcal{T}_A/\omega_0 = 0.2$ and $\mathcal{T}_S/\omega_0 = 0.05$ ($\lambda/\omega_0 = \mathcal{T}_S/\omega_0 = 0.05$).} \label{longcycle}
\end{figure*}

In Fig.~\ref{shortcycle}, we see that in the short cycle limit the machine does not work below a certain value of drive strength $\epsilon$, that depends either on the values of $\mathcal{T}_A$, $\mathcal{T}_S$, for a fixed $\lambda/\omega_0$ ratio (Fig.~\ref{shortcycle}(a)), or the values of $\omega_0$ and $\lambda$, for a fixed $\mathcal{T}_A/\mathcal{T}_S$ ratio (Fig.~\ref{shortcycle}(b)). This can be understood by realising that if $\epsilon$ is too small compared to the incoherent rates $\Gamma_j$, it takes too long to discharge the system and the operational steady state cannot be reached, i.e. the system never stores positive ergotropy. Another way of seeing it is that the linewidths of levels $|S\rangle$ and $|A\rangle$ become much larger than their energy separation, hence they become effectively degenerated in energy. Furthermore, Fig.~\ref{shortcycle}(a) and (b) demonstrate that the efficiency, $\eta$, given by \eqref{geneff}, is always very close or equal to $\eta_{SC}$, given by \eqref{effsc2q}, and that the efficiency increases either when the temperature gradient decreases, for a fixed $\lambda/\omega_0$ ratio, or when the ratio $\lambda/\omega_0$ decreases, for a fixed $\mathcal{T}_A/\mathcal{T}_S$ ratio. Fig.~\ref{shortcycle}(c) shows that the power has the opposite behaviour of the efficiency (Fig.~\ref{shortcycle}(a)) as a function of the temperature gradient, i.e, power increases when the temperature gradient increases. On the other hand, as depicted in Fig.~\ref{shortcycle}(d), the power has the same behaviour of the efficiency (Fig.~\ref{shortcycle}(b)) in terms of the spectrum of the working fluid, i.e, the output power increases when the $\lambda/\omega_0$ ratio decreases.

In Fig.~\ref{longcycle} we display the efficiency and power output for the asymptotic cycle. We see that both quantities have the same qualitative behaviour shown in Fig.~\ref{shortcycle} for the short cycle. However, quantitatively, we see that the efficiency is only close to $\eta_{SC}$, given by \eqref{effsc2q}, when the temperature gradient is sufficiently small (Fig.~\ref{longcycle}(a))  and in Fig.~\ref{longcycle}(b) we see that, for a fixed $\mathcal{T}_A/\mathcal{T}_S$ ratio, the efficiency is always smaller than $\eta_{SC}$.

Looking at the efficiency in quantum thermal machines illustrated in Fig.~\ref{shortcycle}(a) and in Fig.~\ref{longcycle}(a), we observe that in both cases the efficiency decreases when the temperature gradient increases, a phenomenon that was also pointed out in a recent paper by T.~R.~de Oliveira and D.~Jonathan~\cite{Thiago2020} in the different but related context of a four-stroke machine operating on a four level system. Our understanding of this phenomenon is as follows: keeping one temperature fixed, say the hot one, as the temperature gradient increases (the cold temperature is lowered), the Carnot efficiency $\eta^C=1-T_C/T_H$, which represents an upper bound to the engine's efficiency, grows. However, at the same time, an increased thermal gradient pushes the system farther away from equilibrium during the second stroke, that is, it pushes the efficiency $\eta$ farther away below the ideal value $\eta^C$. Our results show that among these two competing effects, the second, namely deviation from ideal operation due to high dissipation, is more important.

\section{Summary}\label{sec:summary}

In this work we have analysed the thermodynamics of a two-stroke quantum heat engine introducing a single qutrit as a working fluid, and comparing its performances to the case of a pair of coupled qubits. In both cases we have studied the machine operating both in the ideal short cycle limit as well as in a cycle with arbitrary duration $\tau$.

For the qutrit, we have seen that the efficiency of the machine does not depend on the duration of cycle but only on the energy spectrum of the working fluid. This happens because work is generated by driving a two-level subspace of the system that can be effectively considered to be in contact with a single negative temperature reservoir. The power output, however, is affected by the duration of the cycle and maximized in the ideal short cycle limit.

For the two coupled qubits as the working fluid we have seen that the efficiency depends on the spectrum of the working fluid, as well as the incoherent transition rates $\Gamma_{A,S}^{k\pm}$, where $k = G,E$, in the ideal short cycle and for a cycle with arbitrary duration $\tau$ depends also on the coupling strength with the external coherent source $\epsilon$. Furthermore, we also saw that the efficiency for cycles with arbitrary duration are less or equal than the efficiency in the ideal short cycle, the latter occurs for long cycles when the temperature gradient is sufficiently small. We also observed that efficiency increases when the temperature difference decreases, a property that was also observed in \cite{Thiago2020}. This was explained based on the fact that at low thermal gradient, the system remains closer to equilibrium, hence it operates closer to the ideal Carnot efficiency, and even though the latter diminishes with decreasing thermal bias, still, the engine efficiency grows.

\section{Acknowledgments}
M.F.S. acknowledges FAPERJ Project No. E-26/202.576/2019 and CNPq Projects No. 302872/2019-1 and INCT-IQ 465469/2014-0. M.F.S. and D.G. would also like to acknowledge the CICOPS program from the University of Pavia for allowing hospitality and support. 
T.F.F.S. acknowledges CAPES for financial support.
M.C. acknowledges financial support from Fondazione CR Firenze Project No. 2018.0951.

\section{Appendix} 

Here we give more details about the results shown in the main text, by analysing the analytics for both the qutrit and the two coupled qubits configurations, respectively.

\subsection{Calculations for the Qutrit in V configuration}

For the qutrit operating in the ideal short cycle limit, i.e, when the discharging stage is adiabatic in the thermodynamic sense and $\sum_j \Gamma_j^{V \pm} \tau \ll 1$, the operational steady state $\rho_{OSS}$ is obtained solving the equation $\rho_{OSS} = \tilde{\rho}_{OSS} + \tau \mathcal{L}(\tilde{\rho}_{OSS})$, where $\tilde{\rho}_{OSS} = U \rho_{OSS} U^{-1}$ and $U = -i (\ket{i}\bra{e} + \ket{e}\bra{i}) + \ket{g}\bra{g} $. In our work, the operational steady state is diagonal in the energy basis, i.e, $\rho_{OSS} = \sum_j  p_j \ket{E_j}\bra{E_j}$, $j = g,e,i$.
 
As described in the main text, the coupling with the reservoirs is given by $\mathcal{L}(\rho) = \sum_k \left[L_k^{V+} + L_k^{V-}\right](\rho)$, $k =e,i$, where $L_k^{V+}(\rho)$ and $L_k^{V-}(\rho)$ are given by Eqs. \eqref{LV+} and \eqref{LV-}, respectively. This interaction defines for the populations of $\rho_{OSS}$ the following equations:

\begin{align}
p_e - p_i &= (\Gamma_e^{V+} p_g - \Gamma_e^{V-} p_i) \tau \\
p_i - p_e &=  (\Gamma_i^{V+} p_g - \Gamma_i^{V-}p_e) \tau \\
p_g &= \frac{\Gamma_i^{V-} p_e + \Gamma_e^{V-} p_i}{\Gamma_i^{V+} + \Gamma_e^{V+}}.
\end{align}
The equations above allow us to calculate $\rho_{oss}$ as a function of the heat exchange rates $\Gamma_k^{V\pm}$ and the cycle duration $\tau$.

After some algebraic manipulation, we obtain

\begin{align}
p_i &= \frac{\Gamma_e^{V+} + \Gamma_i^{V+} - \Gamma_e^{V+} \Gamma_i^{V-} \tau}{\nu } \\
p_e &= \frac{p_i \left[1 - (\Gamma_e^{V-} + \Gamma_e^{V+})\tau\right] + \Gamma_e^{V+} \tau}{1 + \Gamma_e^{V+} \tau},
\end{align}
where
\begin{align}
    \nu &= 2(\Gamma_i^{V+} + \Gamma_e^{V+}) + \Gamma_i^{V-} + \Gamma_e^{V-} \nonumber \\
    &- [\Gamma_i^{V+}(\Gamma_e^{V+} + \Gamma_e^{V-}) + \Gamma_e^{V-} \Gamma_i^{V-} ]\tau.
\end{align}

Using the relations between $p_e$ and $p_i$ and keeping terms up to first order in $\tau$ $\left( \Gamma_j^{V \pm} \tau \ll 1\right)$, the ergotropy, given by $\mathcal{E}_{SC}^{V} = (\omega_i-\omega_e) (p_i - p_e)$, becomes

\begin{align}\label{ergVSC}
\mathcal{E}_{SC}^{V} &= (\omega_i-\omega_e) \tau [(2 \Gamma_e^{V-} + \Gamma_e^{V+}) p_i^V - \Gamma_e^{V+})] \nonumber \\
& = (\omega_i-\omega_e) \tau \frac{\Gamma_i^{V+} \Gamma_e^{V-} - \Gamma_i^{V-} \Gamma_e^{V+}}{2(\Gamma_i^{V+} + \Gamma_e^{V+}) + \Gamma_i^{V-} + \Gamma_e^{V-}}.
\end{align}
Note that, the ergotropy is positive only if $\frac{\Gamma_i^{V+}}{\Gamma_i^{V-}} > \frac{\Gamma_e^{V+}}{\Gamma_e^{V-}}$. It means that for the ergotropy to be positive, effective temperature $T_i^V$ has to be larger than effective temperature $T_e^V$. Using the same approximations and the relation between $p_e$ and $p_i$, the heat exchange with the hot reservoir, $T_i^V$, is given by

\begin{align}\label{qLi}
Q_{SC}^{HV} &= \operatorname{Tr} \left[H_0 (L_i^{V+}(\tilde{\rho}_{OSS})  + L_i^{V-}(\tilde{\rho}_{OSS})) \right]\tau \nonumber \\
&= \omega_i \tau (\Gamma_i^+ p_g - \Gamma_i^- p_e) \nonumber \\
&=\omega_i\tau \left(\Gamma_i^{V+} \frac{\Gamma_i^{V-} p_e + \Gamma_e^{V-} p_i}{\Gamma_i^{V+} + \Gamma_e^{V+}} - \Gamma_i^{V-} p_e \right)\nonumber \\
&=\omega_i \tau \frac{\Gamma_i^{V+} \Gamma_e^{V-} - \Gamma_i^{V-} \Gamma_e^{V+}}{2(\Gamma_i^{V+} + \Gamma_e^{V+}) + \Gamma_i^{V-} + \Gamma_e^{V-}}
\end{align}

From equations \eqref{ergVSC} and \eqref{qLi}, the efficiency of the machine, in the ideal short cycle limit ($SC$), is given by

\begin{equation}\label{effscV}
\eta_{SC}^{V} = \frac{\mathcal{E}_{SC}^{V}}{Q_{SC}^{HV}} = 1- \frac{\omega_e}{\omega_i}.
\end{equation}
Note that, the efficiency does not depend on the rates $\Gamma_k^{V\pm}$ and the maximum efficiency, tending to one, is obtained when $\omega_e \ll \omega_i$.

The power output of the machine, given by $\mathcal{P}_{SC}^{V} = \frac{\mathcal{E}_{SC}^{V}}{\tau}$ in this limit of operation, becomes:

\begin{equation}
\mathcal{P}_{SC}^{V} = (\omega_i-\omega_e) \frac{\Gamma_i^+ \Gamma_e^- - \Gamma_i^- \Gamma_e^+}{2(\Gamma_i^+ + \Gamma_e^+) + \Gamma_g^- + \Gamma_e^-}.
\end{equation}
In the limit of maximum efficiency we have $\mathcal{P}_{SC}^{V} \approx \omega_i \frac{\Gamma_i^+ \Gamma_e^- - \Gamma_i^- \Gamma_e^+}{2(\Gamma_i^+ + \Gamma_e^+) + \Gamma_g^- + \Gamma_e^-}$.

When we consider that the first stage (discharging) of the cycle is no longer adiabatic in the thermodynamic sense, i.e, $\tau_d$ is not short enough, there is heat exchange between the system and the reservoirs. From Eq. \eqref{genw}, the work done by the system is given by

\begin{eqnarray}\label{workV}
  W^V  &=& \int_0^{\tau_d} \operatorname{Tr} \left[\rho(t)\dot{H}(t)\right] dt = \int_0^{\tau_d} i \epsilon (\omega_i-\omega_e) \nonumber \\ &\times& \textrm{Tr}\left[\rho\left(e^{i(\omega_i-\omega_e)t}\ket{e}\bra{i}-e^{-i(\omega_i-\omega_e)t}\ket{i}\bra{e}\right)\right] dt, \nonumber \\
  W^V  &=& \int_0^{\tau_d} i \epsilon (\omega_i-\omega_e) [\varrho_{ie}(t) -\varrho_{ei}(t)] dt, \nonumber \\
\end{eqnarray}
where $\varrho_{jk} = \langle j|e^{iH_0 t} \rho e^{-i H_0 t}|k\rangle$ and $H(t) = H_0 + V^V(t)$. 

The incoming heat from the hot bath in the discharging stage is given by

\begin{align}\label{heatV}
Q_d^{HV} &= \int_0^{\tau_d} \operatorname{Tr} \left[\left[L_i^{V+}(\rho(t)) + L_i^{V-}(\rho(t))\right]H(t)\right] dt \nonumber \\
&= \int_0^{\tau_d} \omega_i \left[\Gamma_i^{V+} \varrho_{gg}(t) - \Gamma_i^{V-} \varrho_{ii}(t)\right] dt \nonumber \\
&-\frac{\epsilon \Gamma_i^{V-}}{2}\int_0^{\tau_d}  \left[\varrho_{ie}(t) + \varrho_{ei}(t)\right] dt.
\end{align}
Note that $\varrho_{kk} = \rho_{kk}$. From Eq. \eqref{eq1}, in the interaction picture, we have

\begin{equation}\label{piV}
\dot{\varrho}_{ii} =  i \epsilon (\varrho_{ie}(t) - \varrho_{ei}) + \Gamma_i^{V+} \varrho_{gg} - \Gamma_i^{V-}\varrho_{ii}
\end{equation}
\begin{equation}\label{peiV}
\dot{\varrho}_{ei} = - i \epsilon (\varrho_{ii} - \varrho_{ee}) - \frac{\Gamma_e^{V-}}{2} \varrho_{ei} - \frac{\Gamma_i^{V-}}{2} \varrho_{ei}
\end{equation}
In the discharging processes $\rho(0) = \rho_{OSS}$. Therefore, from Eq. \eqref{peiV} (and its c.c.) and remembering that $\varrho_{ie}(0) = \varrho_{ei}(0)=0$, it follows that  $ \varrho_{ei}(t) + \varrho_{ie}(t) = 0$. Using this and Eq. \eqref{piV} in Eq. \eqref{heatV}, we obtain

\begin{equation}\label{heatV2}
Q_d^{HV}  = \omega_i \left[\varrho_{ii}(\tau_d) - \varrho_{ii_{OSS}}\right] - \int_0^{\tau_d}  i \epsilon \omega_i \left[\varrho_{ie}(t) -\varrho_{ei}(t)\right] dt.
\end{equation}

In the second stage (recharging), there is only heat exchange between the system and the reservoirs ($V(t) = 0)$ and we consider that we are not restricted to the short cycle limit $\sum_j \Gamma_k^{V\pm} \tau \ll 1$. In this situation, the heat exchanged with the hot reservoir is given by

\begin{align}\label{heatV3}
Q_r^{HV} &= \int_{\tau_d}^{\tau} \operatorname{Tr} \left[\left[L_i^{V+}(\rho(t)) + L_i^{V-}(\rho(t))\right]H_0\right]dt \\ \nonumber 
&= \int_{\tau_d}^{\tau}  \omega_i \left[\Gamma_i^{V+} \varrho_{gg}(t) - \Gamma_i^{V-} \varrho_{ii}(t)\right]dt.
\end{align}
In this stage, we have $\rho(\tau) = \rho_{OSS}$. From Eq. \eqref{eq1}, in the interaction picture, for the recharging stage, we have $\dot{\varrho}_{ii} =  \Gamma_i^{V+} \varrho_{gg} - \Gamma_i^{V-} \varrho_{ii}$, so

\begin{equation}\label{heatV4}
Q_r^{HV} = \omega_i \left[\varrho_{ii_{OSS}} - \varrho_{ii}(\tau_d)\right]
\end{equation}

Finally, the efficiency of the machine is given by

\begin{equation}\label{effV}
\eta_V = \frac{-W^{V}}{Q_d^{HV} + Q_r^{HV}} = 1 - \frac{\omega_e}{\omega_i},
\end{equation}
which is the same as obtained for the short cycle, i.e. for the qutrit in V, the efficiency only depends on its energy levels. Again, maximum efficiency, tending to one, is obtained for $\omega_e \ll \omega_i$.

\subsection{Calculations for the Qutrit in $\Lambda$ configuration}

The system in $\Lambda$ is very similar to the system in $V$, as expected. Once again, the operational steady state in the ideal short cycle is defined by the equation $\rho_{OSS} = \tilde{\rho}_{OSS} + \tau \mathcal{L}(\tilde{\rho}_{OSS})$, where $\tilde{\rho}_{OSS} = U \rho_{OSS} U^{-1}$ but, now, $U = -i (\ket{g}\bra{e} + \ket{e}\bra{g}) + \ket{i}\bra{i}$. The operational steady state is also diagonal in the energy basis: $\rho_{OSS} = \sum_j p_j \ket{E_j}\bra{E_j}$, $j = g, e, i$.

The coupling of the system with the reservoirs is given by $\mathcal{L} (\rho) = \sum_k [L_k^{\Lambda+} + L_k^{\Lambda-}](\rho)$, $k =g,e$, where

\begin{equation}\label{Llambda+}
L_k^{\Lambda+}(\rho) = \Gamma_k^{\Lambda+} \left[\sigma_{i,k} \rho \sigma_{k,i} - \frac{1}{2} \{\sigma_{k,k}, \rho \} \right]
\end{equation}
and
\begin{equation}\label{Llambda-}
L_k^{\Lambda-}(\rho)  = \Gamma_k^{\Lambda-} \left[\sigma_{k,i} \rho \sigma_{i,k} - \frac{1}{2} \{\sigma_{i,i}, \rho \} \right]
\end{equation}

From Eqs. \eqref{Llambda+} and \eqref{Llambda-}, the populations of $\rho_{OSS}$ are defined by the following equations:

\begin{align}
p_g - p_e &= \tau(\Gamma_g^{\Lambda-} p_i - \Gamma_g^{\Lambda+} p_e) \\
p_e - p_g &= \tau (\Gamma_e^{\Lambda-} p_i - \Gamma_e^{\Lambda+}p_g) \\
p_i &= \frac{\Gamma_g^{\Lambda+} p_e + \Gamma_e^{\Lambda+} p_g}{\Gamma_g^{\Lambda-} + \Gamma_e^{\Lambda-}}.
\end{align}
The equations above allow us to calculate $\rho_{OSS}$ in terms of transitions rates $\Gamma_k^{\Lambda \pm}$ and the cycle duration $\tau$.

After some algebraic manipulation, we obtain

\begin{align}
p_e &= \frac{\Gamma_g^{\Lambda-} + \Gamma_e^{\Lambda-} - \Gamma_e^{\Lambda+} \Gamma_g^{\Lambda-} \tau}{\mu} \\
p_g &= \frac{p_e \left[1 - (\Gamma_g^{\Lambda-} + \Gamma_g^{\Lambda+})\tau\right] + \Gamma_g^{\Lambda-} \tau}{1 + \Gamma_g^{\Lambda-} \tau} 
\end{align}
where,

\begin{align}
   \mu &= 2(\Gamma_g^{\Lambda-} + \Gamma_e^{\Lambda-}) + \Gamma_g^{\Lambda+} + \Gamma_e^{\Lambda+} \nonumber \\
   &- [\Gamma_g^{\Lambda+}(\Gamma_e^{\Lambda+} + \Gamma_e^{\Lambda-}) + \Gamma_e^{\Lambda+} \Gamma_g^{\Lambda-} ]\tau
\end{align}
Using the relations between $p_e$ and $p_g$ and keeping terms up to first order in $\tau$ $\left( \Gamma_j^{\Lambda \pm} \tau \ll 1\right)$, the ergotropy, given by $\mathcal{E}_{SC}^{\Lambda} = \omega_e (p_e - p_g)$, becomes

\begin{align}\label{erglambdaSC}
\mathcal{E}_{SC}^{\Lambda} &= \omega_e \tau [(2 \Gamma_g^{\Lambda-} + \Gamma_g^{\Lambda+}) p_e - \Gamma_g^{\Lambda-})] \nonumber \\
& = \omega_e \tau \frac{\Gamma_g^{\Lambda+} \Gamma_e^{\Lambda-} - \Gamma_g^{\Lambda-} \Gamma_e^{\Lambda+}}{2(\Gamma_g^{\Lambda-} + \Gamma_e^{\Lambda-}) + \Gamma_g^{\Lambda+} + \Gamma_e^{\Lambda+}}.
\end{align}
Note that, the ergotropy is positive when $\frac{\Gamma_g^{\Lambda+}}{\Gamma_g^{\Lambda-}} > \frac{\Gamma_e^{\Lambda+}}{\Gamma_e^{\Lambda-}}$. It means that for the ergotropy to be positive the effective temperature $T_g^{\Lambda} = \frac{E_i}{\log{\left[\frac{\Gamma_g^{\Lambda-}}{\Gamma_g^{\Lambda+}}\right]}}$ has to be larger than the effective temperature $T_e^{\Lambda} = \frac{E_i - E_e}{\log{\left[\frac{\Gamma_e^{\Lambda-}}{ \Gamma_e^{\Lambda+}}\right]}}$. Using the same approximations made in \eqref{erglambdaSC} and the relations between $p_e$ and $p_g$, the heat exchange with the hot reservoir is given by

\begin{align}\label{qLg}
Q_{SC}^{H\Lambda} &= \operatorname{Tr} \left[H_0 \left[L_g^{\Lambda+}(\tilde{\rho}_{OSS}) + L_g^{\Lambda-}(\tilde{\rho}_{OSS})\right]  \tau\right] \nonumber \\
&= \omega_i(\Gamma_g^{\Lambda+} p_e - \Gamma_g^{\Lambda-} p_i)\tau \nonumber \\
&=\omega_i\tau \left(\Gamma_g^{\Lambda+} p_e - \Gamma_g^{\Lambda-} \frac{\Gamma_g^{\Lambda+} p_e + \Gamma_e^{\Lambda+} p_g}{\Gamma_g^{\Lambda-} + \Gamma_e^{\Lambda-}} \right)\nonumber \\
&=\omega_i \tau \frac{(\Gamma_g^{\Lambda+} \Gamma_e^{\Lambda-} - \Gamma_g^{\Lambda-} \Gamma_e^{\Lambda+})}{\Gamma_g^{\Lambda-} + \Gamma_e^{\Lambda-}}p_e  =\frac{\omega_i}{\omega_e} \mathcal{E}_{SC}^{\Lambda}
\end{align}

From Eqs. \eqref{erglambdaSC} and \eqref{qLg}, the efficiency of the machine, in the short cycle limit ($SC$), is given by

\begin{equation}
\eta_{SC}^{\Lambda} = \frac{\mathcal{E}_{SC}^{\Lambda}}{Q_{SC}^{H\Lambda}} = \frac{\omega_e}{\omega_i}.
\end{equation}
Note that, as in the V-type configuration, the efficiency does not depend on the transition rates $\Gamma_k^{\Lambda \pm}$, however the maximum efficiency, tending to one, is obtained when $\omega_e \approx \omega_i$.

The power of the machine at this limit of operation, given by $\mathcal{P}_{SC}^{\Lambda} = \frac{\mathcal{E}_{SC}^{\Lambda}}{\tau}$, reduces to

\begin{equation}
\mathcal{P}_{SC}^{\Lambda} = \omega_e  \frac{\Gamma_g^{\Lambda+} \Gamma_e^{\Lambda-} - \Gamma_g^{\Lambda-} \Gamma_e^{\Lambda+}}{2(\Gamma_g^{\Lambda-} + \Gamma_e^{\Lambda-}) + \Gamma_g^{\Lambda+} + \Gamma_e^{\Lambda+}}
\end{equation}
Note that in the limit of maximum efficiency we have $\mathcal{P}_{SC}^{\Lambda} \approx \frac{Q_{SC}^{\Lambda}}{\tau}$.

Similar to the $V$-type configuration, a calculation out of the short cycle limit gives 
\begin{align}\label{worklambda}
W^{\Lambda} &= \int_0^{\tau_d} \operatorname{Tr} \left[\rho(t)\dot{H}(t)\right] dt \nonumber \\
&= \int_0^{\tau_d} i \epsilon \omega_e \left[\varrho_{eg}(t) -\varrho_{ge}(t)\right] dt
\end{align}
for the work extracted from the qutrit, where,  $H(t) = H_0 + V^{\Lambda}(t)$, and

\begin{align}\label{heatlambda}
Q_d^{H\Lambda} &= \int_0^{\tau_d} \operatorname{Tr} \left[\left[L_g^{\Lambda+}(\rho(t)) + L_g^{\Lambda-}(\rho(t))\right]H(t)\right] dt \nonumber \\
&= \int_0^{\tau_d} \omega_i \left[\Gamma_g^{\Lambda+} \varrho_{gg}(t) - \Gamma_g^{\Lambda-} \varrho_{ii}(t)\right] dt \nonumber \\
&-\frac{\epsilon \Gamma_g^{\Lambda+}}{2}\int_0^{\tau_d}  \left[\varrho_{ge}(t) + \varrho_{eg}(t)\right] dt.
\end{align}
as the incoming heat from the hot source in the discharging stage. 

From Eq. \eqref{eq1}, in the interaction picture, we obtain the following equations of motion:

\begin{equation}\label{pglambda}
\dot{\varrho}_{gg} = - i \epsilon (\varrho_{eg} - \varrho_{ge}) - \Gamma_g^{\Lambda+} \varrho_{gg} + \Gamma_g^{\Lambda-} \varrho_{ii}
\end{equation}
\begin{equation}\label{peglambda}
\dot{\varrho}_{eg} = - i \epsilon (\varrho_{gg} - \varrho_{ee}) - \frac{\Gamma_e^{\Lambda+}}{2} \varrho_{eg} - \frac{\Gamma_g^{\Lambda+}}{2} \varrho_{eg}.
\end{equation}
And, given that $\rho(0) = \rho_{OSS}$ and that $\varrho_{ge}(0) = \varrho_{eg}(0)=0$, it follows that $\varrho_{ge}(t) + \varrho_{eg}(t) = 0$ and

\begin{align}\label{heatlambda2}
Q_d^{H\Lambda}  &= \int_0^{\tau_d} i \epsilon \omega_i \left[\varrho_{ge}(t) -\varrho_{eg}(t)\right] dt  \nonumber \\
&- \omega_i \left[\varrho_{gg}(\tau_d) - \varrho_{gg_{OSS}}\right]
\end{align}

In the second stage (recharging), the heat exchanged with the hot reservoir is given by

\begin{align}\label{heatlambda3}
Q_r^{H\Lambda} &= \int_{\tau_d}^{\tau} \operatorname{Tr} \left[\left[L_g^{\Lambda+}(\rho(t)) + L_g^{\Lambda-}(\rho(t))\right]H_0\right] dt \nonumber \\
&= \int_{\tau_d}^{\tau} \omega_i \left[\Gamma_g^{\Lambda+} \varrho_{gg}(t) - \Gamma_g^{\Lambda-} \varrho_{ii}(t)\right] dt.
\end{align}
In this stage we have $\rho(\tau) = \rho_{OSS}$. 

Since $\dot{\varrho}^{\Lambda}_{gg} = - \Gamma_g^{\Lambda +} \varrho_{gg} + \Gamma_g^{\Lambda -} \varrho_{ii}$ we obtain

\begin{equation}\label{heatlambda4}
Q_r^{H\Lambda} = \omega_i \left[\varrho_{gg}(\tau_d) - \varrho_{gg_{OSS}}\right]
\end{equation}

Finally, the efficiency of the machine is given by

\begin{equation}
\eta_\Lambda = \frac{-W^{\Lambda}}{Q_d^{H\Lambda} + Q_r^{H\Lambda}} = \frac{\omega_i}{\omega_e}
\end{equation}

Note that the efficiency depends only on $\omega_e$ and $\omega_i$ as in the short cycle limit and the maximum efficiency, tending to one, is obtained for $\omega_e \approx \omega_i$.

\subsection{Two qubits calculations}

To obtain the operational steady state $\rho_{OSS}$ of the two qubits coupled operating in short cycles we have to solve the equation $\rho_{OSS} =\tilde{\rho}_{OSS} + \tau \mathcal{L}[\tilde{\rho}_{OSS}]$, where $ \tilde{\rho} = U \rho U^{-1}$ and $U = -i\left(\ket{A}\bra{S} + \ket{S}\bra{A} \right) + \ket{G}\bra{G} + \ket{E}\bra{E}$. In the energy eigenbasis $\{\ket{G}, \ket{S}, \ket{A}, \ket{E}\}$, the operational steady state is always diagonal in our work, so $\rho_{OSS}$ is given by $\rho_{OSS} = r_G \ket{G}\bra{G} + r_S \ket{S}\bra{S} + r_A \ket{A}\bra{A} + r_E \ket{E}\bra{E}$.

The populations of $\rho_{OSS}$ are defined by the equations
\begin{equation}\label{ra}
r_A - r_S = \tau [2 \Gamma_A^{E-} r_E + 2 \Gamma_A^{G+} r_G - 2 (\Gamma_A^{E+} + \Gamma_A^{G-}) r_S],
\end{equation}
\begin{equation}\label{rs}
r_S - r_A = \tau[2 \Gamma_S^{G+} r_G + 2 \Gamma_S^{E-} r_E - 2 (\Gamma_S^{E+} + \Gamma_S^{G-}) r_A],
\end{equation}
\begin{equation}\label{rg}
r_G = \frac{\Gamma_S^{G-} r_A + \Gamma_A^{G-} r_S}{\Gamma_A^{G+} + \Gamma_S^{G+}} ,
\end{equation}
\begin{equation}\label{re}
r_E = \frac{\Gamma_A^{E+} r_S + \Gamma_S^{E+} r_A}{\Gamma_S^{E-} + \Gamma_A^{E-}}, 
\end{equation}
where $\Gamma_{A,S}^{k+} = \gamma_{0A,S}^k \bar{n}_{A,S}^k$ and $\Gamma_{A,S}^{k-} = \gamma_{0A,S}^k (\bar{n}_{A,S}^k + 1)$.

Adding equations \eqref{rg} and \eqref{re}, we obtain

\begin{align}\label{rgre}
r_G + r_E &= 1 -(r_A + r_S) =  \frac{\Gamma_S^{G-} r_A + \Gamma_A^{G-} r_S}{\Gamma_A^{G+} + \Gamma_S^{G+}} + \frac{\Gamma_A^{E+} r_S + \Gamma_S^{E+} r_A}{\Gamma_S^{E-} + \Gamma_A^{E-}}   \nonumber \\
1 &= (r_A + r_S) + \frac{\Gamma_S^{G-} r_A + \Gamma_A^{G-} r_S}{\Gamma_A^{G+} + \Gamma_S^{G+}} + \frac{\Gamma_A^{E+} r_S + \Gamma_S^{E+} r_A}{\Gamma_S^{E-} + \Gamma_A^{E-}} \nonumber \\
1 &= r_S\left[1 + \frac{\Gamma_A^{G-}}{\Gamma_A^{G+} + \Gamma_S^{G+}} + \frac{\Gamma_A^{E+}}{\Gamma_S^{E-} + \Gamma_A^{E-}}\right] \nonumber \\
&+ r_A \left[1 + \frac{\Gamma_S^{G-}}{\Gamma_A^{G+} + \Gamma_S^{G+}} + \frac{\Gamma_S^{E+}}{\Gamma_S^{E-} + \Gamma_A^{E-}}\right].
\end{align}

Subtracting equations \eqref{ra} and \eqref{rs}, we have

\begin{align}\label{sub}
r_A - r_S &= \tau [r_G (\Gamma_A^{G+} - \Gamma_S^{G+}) + r_E(\Gamma_A^{E-} - \Gamma_S^{E-}) \nonumber \\
&- r_S (\Gamma_A^{G-} + \Gamma_A^{E+})
+ r_A (\Gamma_S^{E+} + \Gamma_S^{G-}) ]\nonumber \\
r_A - r_S &= \tau\left[ \frac{\Gamma_S^{G-} r_A + \Gamma_A^{G-} r_S}{\Gamma_A^{G+} + \Gamma_S^{G+}} (\Gamma_A^{G+} - \Gamma_S^{G+}) \right. \nonumber \\
&+ \frac{\Gamma_A^{E+} r_S + \Gamma_S^{E+} r_A}{\Gamma_S^{E-} + \Gamma_A^{E-}} (\Gamma_A^{E-} - \Gamma_S^{E-}) \nonumber \\
&- \left.r_S (\Gamma_A^{G-} + \Gamma_A^{E+}) + r_A (\Gamma_S^{E+} + \Gamma_S^{G-})\right].
\end{align}

After some algebraic manipulation the equation above can be written as

\begin{equation}\label{rsra}
r_S = \frac{\alpha}{\beta}r_A,
\end{equation}
where

\begin{align}\label{alpha}
\alpha &= 1 - \tau\left[ \frac{\Gamma_S^{G-}}{\Gamma_G^+} (\Gamma_A^{G+} - \Gamma_S^{G+}) +\frac{\Gamma_S^{E+}}{\Gamma_E^-} (\Gamma_A^{E-} - \Gamma_S^{E-}) \nonumber \right. \\
&+ \Gamma_S^{E+} + \Gamma_S^{G-}],
\end{align}
\begin{align}\label{beta}
\beta &= 1 - \tau\left[\frac{\Gamma_A^{G-}}{\Gamma_G^+} (\Gamma_S^{G+} - \Gamma_A^{G+}) + \frac{\Gamma_A^{E+}}{\Gamma_E^-} (\Gamma_S^{E-} - \Gamma_A^{E-}) \nonumber \right. \\
&+ \Gamma_A^{G-} + \Gamma_A^{E+}]
\end{align}
and $\Gamma_k^{\pm} = \Gamma_A^{k\pm} + \Gamma_S^{k\pm}$ ($k = G, E$).

Using equations \eqref{rsra} and \eqref{rgre}, we obtain the population $r_A$ of the antisymmetric state $\ket{A}$ as a function of $\Gamma_{A,S}^{k\pm}$, $\alpha$ and $\beta$

\begin{equation}
r_A = \frac{\Gamma_G^+ \Gamma_E^-}{\chi},
\end{equation}
where
\begin{align}
    \chi &= (1 + \frac{\alpha}{\beta})\Gamma_G^+ \Gamma_E^- +  \Gamma_E^-(\Gamma_S^{G-} + \frac{\alpha}{\beta}\Gamma_A^{G-}) \nonumber \\ 
    &+ \Gamma_G^+(\Gamma_S^{E+} + \frac{\alpha}{\beta}\Gamma_A^{E+})
\end{align}
Using the relation between $r_A$ and $r_S$ in equation \eqref{rsra} and keeping term up to the first order of $\tau$, the ergotropy, given by $\mathcal{E}_{SC} = 2 \lambda (r_A - r_S)$, becomes

\begin{equation}\label{erg2}
\mathcal{E}_{SC} = \frac{2 \lambda}{\beta}(\beta - \alpha) = 4 \lambda \kappa \tau
\end{equation}
where,

\begin{equation}
    \kappa = \frac{\Gamma_E^{-}(\Gamma_A^{G+} \Gamma_S^{G-} - \Gamma_A^{G-} \Gamma_S^{G+}) \nonumber - \Gamma_G^{+}(\Gamma_A^{E+} \Gamma_S^{E-} - \Gamma_A^{E-} \Gamma_S^{E+})}{\Gamma_E^{-}(\Gamma_G^{+} + \Gamma_G^{-})
    + \Gamma_G^{+}(\Gamma_E^{+} + \Gamma_E^{-})}.
\end{equation}



In this limit of operation, the power of the machine is given by

\begin{align}
\mathcal{P}_{SC} = \frac{\mathcal{E}}{\tau} &= 4 \lambda \kappa
\end{align}


Using the same approximations that we used in the ergotropy, the heat exchange with the hot bath, $Q^A_{SC}$, is given by

\begin{align}\label{qh}
Q^H_{SC} &= \operatorname{Tr}\left[H_0 \mathcal{L}_A (\tilde{\rho}_{OSS}) \tau\right] = \tau [4 \omega_0(\Gamma_A^{E+} r_S - \Gamma_A^{E-} r_E) \nonumber \\
&+ 2 (\omega_0 + \lambda)(\Gamma_A^{G+} r_G - \Gamma_A^{E+} r_S + \Gamma_A^{E-} r_E - \Gamma_A^{G-} r_S)]  \nonumber \\
&=  \frac{2 (\omega_0 - \lambda) \tau}{\Gamma_S^{E-} + \Gamma_A^{E-}} (\Gamma_A^{E+}\Gamma_S^{E-} - \Gamma_S^{E+}\Gamma_A^{E-})r_A\nonumber \\
&+ \frac{2 (\omega_0 + \lambda) \tau}{\Gamma_S^{G+} + \Gamma_A^{G+}} (\Gamma_A^{G+}\Gamma_S^{G-} - \Gamma_S^{G+}\Gamma_A^{G-})r_A  \nonumber \\
&= \frac{2 \tau}{\Omega} [(\omega_0 - \lambda)\Gamma_G^+(\Gamma_A^{E+}\Gamma_S^{E-} - \Gamma_S^{E+}\Gamma_A^{E-}) \nonumber\\
&+ (\omega_0 + \lambda)\Gamma_E^- (\Gamma_A^{G+}\Gamma_S^{G-} - \Gamma_S^{G+}\Gamma_A^{G-})], \nonumber \\
\end{align}
where
\begin{equation}
    \Omega = \Gamma_E^{-}(\Gamma_G^{+} + \Gamma_G^{-})
    + \Gamma_G^{+}(\Gamma_E^{+} + \Gamma_E^{-}). \nonumber
\end{equation}
Finally, we calculate the efficiency of the machine operating in the short cycle limit and we obtain

\begin{equation}
    \eta_{SC}  = \left(1 - \frac{E_S}{E_A}\right) \frac{1 - f}{1 + \frac{E_S}{E_A} f}
\end{equation}
where,

\begin{equation}
    f = \frac{\Gamma_G^+(\Gamma_A^{E+}\Gamma_S^{E-} - \Gamma_S^{E+}\Gamma_A^{E-})}{\Gamma_E^- (\Gamma_A^{G+}\Gamma_S^{G-} - \Gamma_S^{G+}\Gamma_A^{G-})}.
\end{equation}


Maximum efficiency is obtained when $f << 1$. This condition is reached when $\Gamma_A^{k-} + \Gamma_S^{k-} >> \Gamma_A^{k+} + \Gamma_S^{k+}$. This can be achieved either if $\Gamma_j^{k-} >> \Gamma_j^{k+}$ or if $\Gamma_S^{k-} >> \Gamma_A^{k-}$. In both cases, the maximum efficiency tends to
\begin{equation}\label{efmaxsc}
    \eta_{SC}^{max} \rightarrow 1 - \frac{E_S}{E_A}
\end{equation}
Note that the efficiency tending to one, when $\lambda \approx \omega_0$.

For $\Gamma_j^{k-} >> \Gamma_j^{k+}$, the power in the short cycle depends on both temperature and is given by $\mathcal{P}_{SC}^{max} \approx 4 \lambda (\Gamma_A^{G+}\Gamma_S^{G-} - \Gamma_S^{G+}\Gamma_A^{G-})/\Gamma_G^-$. For $\Gamma_S^{k-} >> \Gamma_A^{k-}$ the power output depends only the temperature of the hot reservoir and is given by $\mathcal{P}_{SC}^{max} \approx 4 \lambda \Gamma_A^{G+}$.


When the discharging stage is not adiabatic in the thermodynamic sense, the heat exchange with the hot bath in this stage is given by

\begin{align}\label{heat2cq}
    Q_d^H & = \int_0^{\tau_d} \operatorname{Tr} \left[\mathcal{L}_A(\rho(t)) H(t)\right]dt \nonumber \\
    &= \int_0^{\tau_d} \{4\omega_0\left[\Gamma_A^{E+} \varrho_A(t) - \Gamma_A^{E-} \varrho_{E}(t)\right] \nonumber \\
&+ (\omega_0 + \lambda) [ 2 \Gamma_A^{G+} \varrho_{G}(t) + 2 \Gamma_A^{E-}  \varrho_{E}(t) \nonumber \\
&- 2 (\Gamma_A^{G-} + \Gamma_A^{E+}) \varrho_A(t)]\} dt \nonumber \\
&- \epsilon (\Gamma_A^{E+} +  \Gamma_S^{E+} + \Gamma_S^{G-} + \Gamma_A^{G-}) \int_0^{\tau_d} (\varrho_{AS}(t) + \varrho_{SA}(t))dt.
\end{align}

By equation \eqref{genw} the work done by the system in this stage is given by

\begin{align} \label{work2cq}
W &= \int_0^{\tau_d} \operatorname{Tr}\left[\rho(t) \dot{H}(t)\right] dt \nonumber \\
&= 2i \epsilon \lambda \int_0^{\tau_d} \left[\varrho_{AS}(t) - \varrho_{SA}(t)\right] dt,
\end{align}
where $H(t) = H_0 + V(t)$.

The equations of motion of the two coupled qubtis, in the interaction picture, in the discharging process are given by



\begin{align}
\dot{\varrho}_{E} &= 2 \Gamma_S^{E+} \varrho_S + 2 \Gamma_A^{E+} \varrho_A - 2 (\Gamma_S^{e-} + \Gamma_A^{e-}) \varrho_{E}, \\
\dot{\varrho}_{G} &= 2 \Gamma_S^{G-} \varrho_S + 2 \Gamma_A^{G-} \varrho_A - 2 (\Gamma_S^{G+} + \Gamma_A^{G+}) \varrho_{G}, \\\label{pad}
\dot{\varrho}_{A} & = -i \epsilon (\varrho_{SA}-\varrho_{AS}) + 2 \Gamma_A^{G+} \varrho_{gg} + 2 \Gamma_A^{E-}  \varrho_{E} \nonumber \\
&- 2 (\Gamma_A^{G-} + \Gamma_A^{E+}) \varrho_A, \\
\dot{\varrho}_{S} & = i \epsilon (\varrho_{SA}-\varrho_{AS}) + 2 \Gamma_S^{G+} \varrho_{G} + 2 \Gamma_S^{E-}  \varrho_{E}  \nonumber \\
&- 2 (\Gamma_S^{G-} + \Gamma_S^{E+}) \varrho_S, \\
\dot{\varrho}_{SA} & = -i \epsilon (\varrho_{A}-\varrho_{S}) - (\Gamma_A^{E+} +  \Gamma_S^{E+} + \Gamma_S^{G-} + \Gamma_A^{G-}) \varrho_{SA}, \\
\dot{\varrho}_{AS} & = i \epsilon (\varrho_{A}-\varrho_{S}) - (\Gamma_A^{E+} + \Gamma_S^{E+} + \Gamma_S^{G-} + \Gamma_A^{G-}) \varrho_{AS}.
\end{align}



Substituting Eq.\eqref{pad} in Eq. \eqref{work2cq}, we obtain

\begin{eqnarray}\label{work2cq2}
W &=& 2 \lambda [\varrho_A(\tau_d) -\varrho_{A_{OSS}} \nonumber \\
&+& 2 (\Gamma_A^{G-} + \Gamma_A^{E+}) \int_0^{\tau_d} \varrho_A(t)  dt -  2 \Gamma_A^{G+} \int_0^{\tau_d} \varrho_{G}(t) dt \nonumber \\
&-&2 \Gamma_A^{E-} \int_0^{\tau_d} \varrho_{E}(t) dt ]
\end{eqnarray}
where, $\rho_A(0) = \rho_{A_{OSS}}$. When $\epsilon \gg \Gamma_{A,S}^{E \pm}, \Gamma_{A,S}^{G \pm}$ the swap between the populations is approximately instantaneous ($\tau_d \rightarrow 0$). In that situation, by Eq. \eqref{work2cq2}, the work is given by

\begin{equation}
    W = 2 \lambda (\varrho_{S_{OSS}} - \varrho_{A_{OSS}}) = 2 \lambda (r_S - r_A) = - \mathcal{E}
\end{equation}
where $\mathcal{E}$ is the ergotropy stored in the operational steady $\rho_{OSS}$.


Substituting \eqref{pad} in \eqref{heat2cq}, we obtain
\begin{align}
Q_d^H & =  \int_0^{\tau_d} 4 \omega_0\left[\Gamma_A^{E+} \varrho_A(t) - \Gamma_A^{E-} \varrho_{E}(t)\right] dt  \nonumber \\
&+ i \epsilon (\omega_0 + \lambda)\int_0^{\tau_d} \left[\varrho_{SA}(t)-\varrho_{AS}(t)\right]dt \nonumber \\
&- (\lambda + \omega_0) [\varrho_{A_{OSS}} - \varrho_A(\tau_d)].
\end{align}
In $t = 0$ the system is in the operational steady state $\rho_{OSS}$ that is diagonal in the energy basis, so by the equations of motions for the coherence we see that $\varrho_{AS}(t) + \varrho_{SA}(t) = 0$. Under the condition where the swap between the populations is approximately instantaneous ($\tau_d \rightarrow 0$, $\epsilon >> \Gamma_{A,S}^{E \pm}, \Gamma_{A,S}^{E \pm}$) we see that $Q_d^H \rightarrow 0$. It means that the process is adiabatic in the thermodynamic sense in this limit, as expected.

In the recharging process there is no external field ($\epsilon = 0$), so $W = 0$. Now we consider that we are not restricted to the short cycle limit, so the heat exchanged with the hot bath is given by
\begin{align}
Q_r^H &= \int_0^{\tau_d} \operatorname{Tr} \left[\mathcal{L}_A(\rho(t)) H_0\right]dt \nonumber \\
&=\int_{\tau_d}^{\tau} \{4 \omega_0\left[\Gamma_A^{E+} \varrho_A(t) - \Gamma_A^{E-} \varrho_{E}(t) \right] \nonumber\\
&+ (\omega_0 + \lambda) [ 2 \Gamma_A^{G+} \varrho_{G} + 2 \Gamma_A^{E-}  \varrho_{E}(t) \nonumber \\
&- 2 (\Gamma_A^{G-} + \Gamma_A^{E+}) \varrho_A(t)]\} dt \nonumber \\
 &= \int_{\tau_d}^{\tau} 4\omega_0\left[\Gamma_A^{E+} \varrho_A(t) - \Gamma_A^{E-} \varrho_{E}(t) \right]dt \nonumber \\
 &+ (\omega_0 + \lambda)[\varrho_A(\tau) - \varrho_A(\tau_d)] \nonumber \\
 &= \int_{\tau_d}^{\tau} 4 \omega_0 \left[\Gamma_A^{E+} \varrho_A(t) - \Gamma_A^{E-} \varrho_{E}(t) \right]dt \nonumber \\
 &+ (\omega + \lambda)[\varrho_{A_{OSS}} - \varrho_A(\tau_d)].
\end{align}
In this stage the system back to the initial state $\rho_{OSS}$, so $\rho_A(\tau) = \rho_{A_{OSS}}$.

Finally, the efficiency of the machine is given by

\begin{equation}
\eta = \frac{-W}{Q_d^H + Q_r^H} = \frac{1 - \frac{E_S}{E_A}}{1 + \alpha + \beta},
\end{equation}
where,

\begin{align}
\alpha &= \frac{\int_0^{\tau_d}  4 \omega_0\left[\Gamma_A^{E+} \varrho_A(t) - \Gamma_A^{E-} \varrho_{E}(t)\right] dt }{i\epsilon (\omega_0 + \lambda) \int_0^{\tau_d} \left[\varrho_{SA}(t)-\varrho_{AS}(t)\right] dt} \\
\beta &= \frac{\int_{\tau_d}^{\tau} 4 \omega_0\left[\Gamma_A^{E+} \varrho_A(t) - \Gamma_A^{E-} \varrho_{E}(t)\right] dt}{i\epsilon (\omega_0 + \lambda) \int_0^{\tau_d} \left[\varrho_{SA}(t)-\varrho_{AS}(t)\right] dt} .
\end{align}
Different from the qutrit case, here we see that the efficiency of the machine depends on $\epsilon$. 

For a large enough  $\tau$ the operational steady state, $\rho_{OSS}$, converges to the {\it bona fide} steady state, $\rho_{NESS}$, defined by $\dot{\rho} = 0$. In the steady state  $\rho_{NESS}$ the populations are defined by the following equations:

\begin{equation}\label{ress}
r_{E_{NESS}} = \frac{\Gamma_S^{E+} r_{S_{NESS}}  + \Gamma_A^{E+} r_{A_{NESS}}  }{\Gamma_S^{E-} + \Gamma_A^{E-}}
\end{equation}
\begin{equation}\label{rgss}
r_{G_{NESS}}  = \frac{\Gamma_S^{G-} r_{S_{NESS}}  + \Gamma_A^{G-} r_{A_{NESS}}  }{\Gamma_S^{G+} + \Gamma_A^{G+}}
\end{equation}
\begin{equation}\label{rass}
r_{A_{NESS}}  = \frac{\Gamma_A^{G+} r_{G_{NESS}}  + \Gamma_A^{E-} r_{E_{NESS}}}{\Gamma_A^{E+} + \Gamma_A^{G-}}
\end{equation}
\begin{equation}\label{rsss}
r_{S_{NESS}}  = \frac{\Gamma_S^{g+} r_{G_{NESS}}  + \Gamma_g^{E-} r_{e_{NESS}}  }{\Gamma_S^{E+} + \Gamma_S^{G-}}.
\end{equation}

Adding equations \eqref{ress} e \eqref{rgss}, we obtain

\begin{align}\label{um}
r_{G_{NESS}}  + r_{E_{NESS}}  &= 1 - (r_{A_{NESS}}  + r_{S_{NESS}} ) \nonumber \\
&=  \frac{\Gamma_S^{G-} r_{S_{NESS}}  + \Gamma_A^{G-} r_{A_{NESS}}  }{\Gamma_G^+} \nonumber \\
&+ \frac{\Gamma_S^{E+} r_{S_{NESS}}  + \Gamma_A^{E+} r_{A_{NESS}}  }{\Gamma_E^-} \nonumber \\
1 &= r_{A_{NESS}}  \left[1 + \frac{\Gamma_A^{G-}}{\Gamma_G^+} +  \frac{\Gamma_A^{E+}}{\Gamma_E^-}\right] \nonumber \\
&+ r_{S_{NESS}}  \left[1 + \frac{\Gamma_S^{G-}}{\Gamma_G^+} +  \frac{\Gamma_S^{E+}}{\Gamma_E^-}\right] \nonumber \\
&= r_{A_{NESS}}  \nu + r_{S_{NESS}}  \mu
\end{align}
where,

\begin{equation}\label{nu}
\nu = 1 + \frac{\Gamma_A^{G-}}{\Gamma_G^+} +  \frac{\Gamma_A^{E+}}{\Gamma_E^-}
\end{equation}

\begin{equation} \label{mu}
\mu = 1 + \frac{\Gamma_S^{G-}}{\Gamma_G^+} +  \frac{\Gamma_S^{E+}}{\Gamma_E^-}
\end{equation}

By equation \eqref{rass}, we have

\begin{align}
(\Gamma_A^{E+} + \Gamma_A^{G-} )r_{A_{NESS}} &= \Gamma_A^{G+} \frac{\Gamma_S^{G-} r_{S_{NESS}} + \Gamma_A^{G-} r_{A_{NESS}} }{\Gamma_G^+} \nonumber \\
&+ \Gamma_A^{E-}  \frac{\Gamma_S^{E+} r_{S_{NESS}} + \Gamma_A^{E+} r_{A_{NESS}} }{\Gamma_E^-}.
\end{align}
Multiplying both sides by $\Gamma_E^-\Gamma_G^+$, we obtain
\begin{align}
&\Gamma_E^-\Gamma_G^+(\Gamma_A^{E+} + \Gamma_A^{G-} )r_{A_{NESS}} \nonumber \\
&= \Gamma_A^{G+} \Gamma_E^- (\Gamma_S^{G-} r_{S_{NESS}} + \Gamma_A^{G-}  r_{A_{NESS}}) \nonumber \\
&+ \Gamma_A^{E-} \Gamma_G^+ (\Gamma_S^{E+} r_{S_{NESS}} + \Gamma_A^{E+} r_{A_{NESS}}).
\end{align}
After some algebraic manipulation, we obtain

\begin{equation}\label{rsrass}
r_{S_{NESS}} = \frac{\epsilon}{\zeta} r_{A_{NESS}}
\end{equation}
where,

\begin{equation}\label{epsilon}
\epsilon = \Gamma_A^{E+} \Gamma_S^{E-}\Gamma_G^+ + \Gamma_S^{G+} \Gamma_A^{G-} \Gamma_E^-
\end{equation}

\begin{equation}\label{zeta}
\zeta = \Gamma_A^{G+} \Gamma_S^{G-} \Gamma_E^- + \Gamma_A^{E-} \Gamma_S^{E+} \Gamma_G^+
\end{equation}
By equations \eqref{rsrass} and \eqref{um}, we have

\begin{equation}\label{rasol}
r_{A_{NESS}} = \frac{\zeta}{K}
\end{equation}
where,

\begin{align}\label{k}
  K &= \Gamma_A^{E+} \Gamma_S^{G-} (\Gamma_S^{E-} + \Gamma_A^{G+}) + \Gamma_S^{E+} \Gamma_A^{G-} (\Gamma_A^{E-} + \Gamma_S^{G+}) \nonumber\\
  &+ \Gamma_E^- [\Gamma_A^{G-} \Gamma_S^{G+} + \Gamma_S^{G-}(\Gamma_A^{G-} + \Gamma_A^{G+})] \nonumber \\
&+ \Gamma_G^+ [\Gamma_A^{E-} \Gamma_S^{E+} + \Gamma_A^{E+}(\Gamma_S^{E+} + \Gamma_S^{E-})].
\end{align}

By equation \eqref{rasol} and \eqref{rsrass}, we have

\begin{equation}\label{rssol}
r_{S_{NESS}} = \frac{\epsilon}{K}.
\end{equation}

\begin{figure}
  \centering
   \includegraphics[width=\columnwidth]{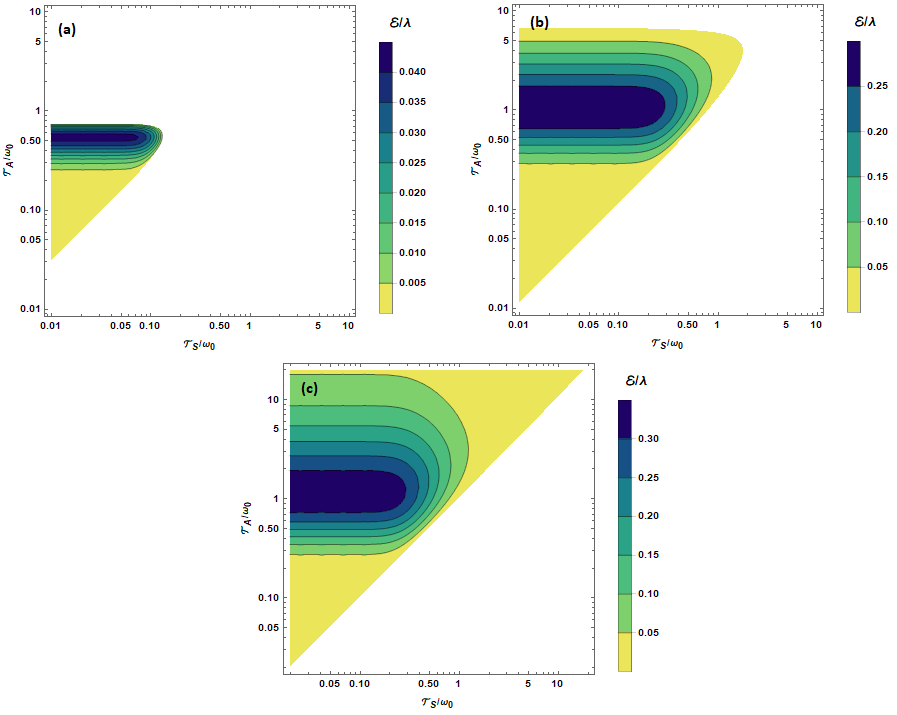}
   \caption{We plot $\mathcal{E}_{NESS}/ \lambda$ as a function $\mathcal{T}_A$ and $\mathcal{T}_S$ for $\lambda / \omega_0 = 0.5$ (Fig. (a)), $\lambda/\omega_0 = 0.05$ (Fig. (b)) and $\lambda/ \omega_0 = 0.001$ (Fig. (c)). The vertical bars indicate the value of $\mathcal{E}_{NESS}/ \lambda$ in each colored region.} \label{ergotropyness}
\end{figure}

Using \eqref{rasol} and \eqref{rssol} the ergotropy, given by $\mathcal{E}_{NESS} = 2 \lambda (r_{A_{NESS}} - r_{S_{NESS}})$, becomes

\begin{align}\label{ergss}
\mathcal{E}_{NESS} &= \frac{2 \lambda}{K}(\zeta - \epsilon) \nonumber \\
&=\frac{2 \lambda}{K} [\Gamma_E^-(\Gamma_A^{G+} \Gamma_S^{G-} -\Gamma_S^{G+} \Gamma_A^{G-} ) \nonumber \\
&- \Gamma_G^+(\Gamma_A^{E+} \Gamma_S^{E-}  - \Gamma_A^{E-} \Gamma_S^{E+})]
\end{align}
In fig. \eqref{ergotropyness} we give a numerical example for the ergotropy in the steady state, $\mathcal{E}_{NESS}$, as a function of $\mathcal{T}_A/\omega_0$ and $\mathcal{T}_S/\omega_0$ for different values of $\lambda/\omega_0$. In Fig. \eqref{ergotropyness} (c) we see that when $\lambda \ll \omega_0$, whenever there is a positive temperature gradient from $\mathcal{T}_A$ to $\mathcal{T}_S$ we have the necessary conditions to store ergotropy in the system. In Fig. \eqref{ergotropyness} (a) and Fig. \eqref{ergotropyness} (b), we see that a positive temperature gradient from $\mathcal{T}_A$ to $\mathcal{T}_S$ is not a sufficient condition to store ergotropy.
 
\end{document}